\documentclass[a4paper,11pt]{article}

\usepackage[margin=25mm]{geometry}
\usepackage[T1]{fontenc}
\usepackage[utf8]{inputenc}
\usepackage{lmodern}
\usepackage[english]{babel}
\usepackage{amsmath,amssymb,mathtools,bm,amsthm}
\usepackage{graphicx}
\usepackage{placeins,flafter}
\usepackage{tikz}
\usetikzlibrary{arrows.meta,calc,positioning}
\usepackage{float}
\usepackage{xcolor,microtype,etoolbox,needspace}
\usepackage[numbers,sort&compress]{natbib}
\usepackage{xurl,hyperref}
\usepackage{caption}
\definecolor{linkblue}{HTML}{005C99}
\definecolor{citegreen}{HTML}{087F5B}
\definecolor{edgeblue}{HTML}{0072B2}
\hypersetup{colorlinks=true,linkcolor=linkblue,citecolor=citegreen,
  urlcolor=linkblue,pdfauthor={Andrey Zhukov},
  pdfkeywords={Heisenberg model, Dzyaloshinskii-Moriya interaction, Hamiltonian simulation,
  quantum circuits, product formulas, chiral spin transport, spin triangle}}
\numberwithin{equation}{section}
\newcommand{\ii}{\mathrm{i}}
\newcommand{\Id}{\mathbb I}

\newcommand{\Ugate}[2]{U_{#1}\!\left(#2\right)}
\newcommand{\atanTwo}{\operatorname{atan2}}
\newcommand{\Tr}{\operatorname{Tr}}
\newcommand{\kagomeTriangle}[2]{\vcenter{\hbox{\rotatebox[origin=c]{#2}{$#1\triangle$}}}}
\newcommand{\triup}{\mathord{\mathpalette\kagomeTriangle{0}}}
\newcommand{\tridown}{\mathord{\mathpalette\kagomeTriangle{180}}}

\newcommand{\AuthorName}{Andrey Zhukov}
\newcommand{\Affiliation}{}
\newcommand{\AuthorEmail}{zugazoid@gmail.com}

\newcommand{\DatasetVersionDOI}{10.5281/zenodo.23007728}

\author{\AuthorName\ifdefempty{\AuthorEmail}{}{\thanks{\href{mailto:\AuthorEmail}{\AuthorEmail}}}\ifdefempty{\Affiliation}{}{\\[3pt]{\small\Affiliation}}}
\date{}

\newenvironment{revisionblock}{}{}

\title{\bfseries Exact quantum circuits for a Heisenberg triangle
with Dzyaloshinskii--Moriya interaction}
\hypersetup{pdftitle={Exact quantum circuits for a Heisenberg triangle with Dzyaloshinskii-Moriya interaction}}
\begin{document}
\maketitle

\begin{abstract}
We construct short analytical quantum circuits for the exact evolution
of a spin triangle with Dzyaloshinskii--Moriya (DM) interaction.
We consider pure DM interaction with arbitrary real couplings and two
Hamiltonian families with uniform Heisenberg exchange. For pure DM,
a change of basis reduces the evolution to two single-qubit rotations.
The circuits reproduce the full evolution operator at arbitrary time
with a fixed gate count and admit implementations with both triangular
and linear qubit connectivity. Experiments on a superconducting IBM
quantum processor demonstrate directional quantum-state transfer and
reproduce the amplitude and phase of transverse magnetization.
In these experiments, the exact circuits reduce distortions
relative to circuits based on the second-order Trotter formula.
On a 12-spin kagome cluster, exact triangle blocks reduce the gate count
at a prescribed accuracy. We also investigate the reduction of coherent
errors by adjusting circuit angles and applying echo pulses to neighboring
qubits.
\end{abstract}


\section{Introduction}

Frustrated quantum magnets provide a setting for studying how competing
interactions produce collective states and excitations. Connecting their
microscopic models to experiments requires calculations of the dynamical
spin correlations that determine magnetic excitation spectra. Neutron
scattering on the kagome antiferromagnet herbertsmithite, for example,
has revealed a broad excitation continuum interpreted as evidence of
fractionalized spin excitations \cite{Han2012Fractionalized}.
Such measurements motivate the development of methods for computing
spin dynamics and distinguishing competing descriptions of magnetic materials.

The Dzyaloshinskii--Moriya (DM) interaction is an antisymmetric exchange
arising from spin--orbit coupling when bond symmetry permits it
\cite{Dzyaloshinsky1958Weak,Moriya1960Anisotropic}.
In kagome magnets, it can modify the dynamical response and favor
magnetic order \cite{Prelovsek2021Dynamics}.
The triangle is a natural local unit of the kagome lattice and the
smallest closed loop with competing antiferromagnetic bonds.
Chiral spin dynamics has been implemented in nuclear magnetic resonance
and superconducting circuits \cite{Lu2016Chiral,Wang2019Antisymmetric}.

Quantum processors offer a route to dynamical correlations and the
associated spectroscopic observables \cite{Chiesa2019Neutron}.
Digital simulation commonly alternates bond evolutions using product
formulas \cite{Trotter1959Product,Strang1968Splitting}.
Our main reference approximation is the symmetric second-order Trotter
formula, also called Strang splitting. Reducing the error due to
noncommuting bonds generally requires more gates
\cite{Childs2021Trotter}.
Another route to shorter circuits exploits the structure of the Hamiltonian.
For particular model classes, Cartan decomposition gives exact fixed-depth
circuits \cite{Kokcu2022Cartan}, while algebraic compression of gate
sequences has been developed for free-fermion models
\cite{Kokcu2022AlgebraicCompression,Camps2022CompressionAlgorithm},
including extensions to periodic spin chains \cite{Kokcu2025Periodic}.

For lattices of coupled triangles, circuits implementing the evolution
of an entire three-spin cluster are of particular interest.
Triangle blocks have been used in kagome simulation with collective
three-spin operations \cite{Maskara2023Programmable}.
Exact decompositions for Heisenberg exchange and three-spin scalar
chirality have been obtained using local symmetries \cite{NegishiYang2026}.
Such solutions incorporate the interactions within a block without
the error of splitting them into separate bonds. Finding short exact
circuits for other physically relevant triangle Hamiltonians expands
the set of local operations available for simulating spin lattices.

Here we construct short circuits with explicit analytical parameters
for triangles with two-spin DM interactions. We consider pure DM with
arbitrary real couplings, equal DM couplings with uniform Heisenberg
exchange, and a phase-constrained family of unequal DM couplings with
uniform exchange. A change of basis gives the full evolution operator
at arbitrary time with a fixed gate count. The same families admit
implementations on a qubit chain without SWAP gates.
For pure DM, analytical angle choices turn the five-DM-gate sequence
used in a Strang step into an exact decomposition.

On a superconducting IBM quantum processor, we compare the exact
circuits with bondwise Strang splitting through excitation transfer, the amplitude and phase of
transverse magnetization, and single-qubit state transfer. A 12-spin kagome calculation then assesses
the cost of using exact triangles as local blocks at a prescribed
accuracy; the combination of neighboring blocks remains approximate.
A separate numerical demonstration considers reducing a known coherent
ZZ error by adjusting existing angles and protecting inactive neighbors
when composing the blocks.
Appendix~\ref{app:circuits} provides the circuits, parameters and short
derivations, while the Supplemental Material gives further proofs,
resource analyses and numerical details.

\Needspace{13\baselineskip}
\section{Exact circuits from a change of basis}
\label{sec:exact}

Consider three spin-$1/2$ particles with Hamiltonian
\begin{equation}
H=H_J+H_D
=\sum_{(ij)\in\{12,23,31\}}\Bigl[
J(X_iX_j+Y_iY_j+Z_iZ_j)
+D_{ij}(X_iY_j-Y_iX_j)
\Bigr].
\label{eq:Hfull}
\end{equation}
Here $X_i,Y_i,Z_i$ are Pauli operators acting on particle $i$;
$H_J$ is the isotropic exchange term with uniform coupling $J$, and
$H_D$ is the DM term with arbitrary real couplings $D_{ij}$.
All DM vectors lie along a common $z$ axis; $D_{ij}$ are their signed coefficients.
In all three-spin constructions below, the exchange constant $J$ is the same
on all three bonds.
The sum runs over the oriented edges $12,23,31$ (Fig.~\ref{fig:triangle});
$\hbar=1$. In the spin-operator convention $\mathbf{S}_i=(X_i,Y_i,Z_i)/2$,
the corresponding couplings are $4J$ and $4D_{ij}$.

We consider three families: pure DM with arbitrary couplings,
equal DM couplings at arbitrary $J$, and unequal couplings with nonzero
exchange subject to Eq.~\eqref{eq:fluxsurface}. These constructions do
not constitute a classification of all exactly implementable three-spin
Hamiltonians.

This Hamiltonian conserves the excitation number, which counts the spins
in state $|1\rangle$. The corresponding operator is
$N=\sum_i(\Id-Z_i)/2$, where $\Id$ denotes the identity.

\begin{figure}[H]
\centering
\begin{tikzpicture}[scale=.83,>=Latex]
\node[circle,draw,fill=white,minimum size=8mm] (q1) at (0.000,2.944) {$1$};
\node[circle,draw,fill=white,minimum size=8mm] (q2) at (-1.700,0.000) {$2$};
\node[circle,draw,fill=white,minimum size=8mm] (q3) at (1.700,0.000) {$3$};
\draw[->,line width=1.2pt,edgeblue] (q1)--
  node[midway,left=3pt,font=\small] {$D_{12}$} (q2);
\draw[->,line width=1.2pt,edgeblue] (q2)--
  node[midway,below=3pt,font=\small] {$D_{23}$} (q3);
\draw[->,line width=1.2pt,edgeblue] (q3)--
  node[midway,right=3pt,font=\small] {$D_{31}$} (q1);
\end{tikzpicture}
\caption{DM bond orientation: $1\to2\to3\to1$.
Each edge also carries exchange $J$. Reversing an arrow changes the
sign of the corresponding $D_{ij}$.}
\label{fig:triangle}
\end{figure}

\subsection{Pure DM}

At $J=0$, the Hamiltonian~\eqref{eq:Hfull} reduces to the DM interaction,
\begin{equation}
H_D=\sum_{(ij)\in\{12,23,31\}}D_{ij}(X_iY_j-Y_iX_j).
\label{eq:pure-dm}
\end{equation}
We construct a unitary transformation
$W(D_{12},D_{23},D_{31})$ such that
\begin{equation}
WH_DW^\dagger=\Omega(Z_1+Z_2),\qquad
\Omega=\sqrt{D_{12}^2+D_{23}^2+D_{31}^2}.
\label{eq:encoder-diagonal}
\end{equation}
In this basis, the Hamiltonian contains only two independent local fields.
The third qubit participates in $W$ but is idle during the central evolution.

\begin{figure}[!ht]
\centering
\includegraphics[width=\linewidth]{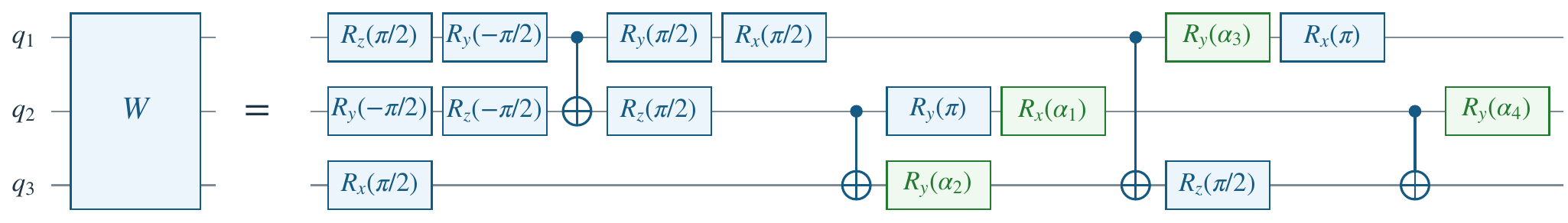}
\caption{Explicit basis change $W$ for arbitrary pure DM couplings.
Green gates have coupling-dependent angles $\alpha_i$;
blue gates are fixed.}
\label{fig:encoder-four}
\end{figure}

The circuit in Fig.~\ref{fig:encoder-four} defines $W$ explicitly.
It consists of 4~CNOT gates and single-qubit rotations
$R_k(\theta)=e^{-\ii\theta\sigma_k/2}$, $k\in\{x,y,z\}$. The four angles
$\alpha_1,\ldots,\alpha_4$ depend only on the couplings $D_{ij}$;
their explicit expressions are given in Eq.~\eqref{eq:encoder-angles}
of Appendix~\ref{app:circuits}.

Evolution in this basis reduces to two single-qubit rotations:
\begin{equation}
e^{-\ii tH_D}=W^\dagger
R_z^{(1)}(2\Omega t)R_z^{(2)}(2\Omega t)W.
\label{eq:encoder-dm}
\end{equation}
The transformations $W$ and $W^\dagger$ together require
at most 8~CNOT gates.
The evolution circuit is shown in Fig.~\ref{fig:diagonal-circuits}(a).
These CNOT counts assume unrestricted pair connectivity.
Section~\ref{sec:path} gives the corresponding circuits for the
nearest-neighbor chain $1$--$2$--$3$.

\Needspace{14\baselineskip}
\subsection{Heisenberg--DM with uniform couplings}
\label{sec:uniform}

Now consider the Hamiltonian~\eqref{eq:Hfull} with uniform DM couplings,
$D_{12}=D_{23}=D_{31}=D$, and arbitrary exchange $J$.
Write the Hamiltonian in terms of two dimensionless operators:
\begin{equation}
\begin{gathered}
H=JK+DH_0,\\
K=\sum_{(ij)\in\{12,23,31\}}(X_iX_j+Y_iY_j+Z_iZ_j),\\
H_0=\sum_{(ij)\in\{12,23,31\}}(X_iY_j-Y_iX_j).
\end{gathered}
\label{eq:uniform-model}
\end{equation}
The DM operator $H_0$ is diagonalized by the previously constructed
transformation $W_0=W(1,1,1)$: setting $D_{12}=D_{23}=D_{31}=1$
in Eq.~\eqref{eq:encoder-diagonal} gives
$W_0H_0W_0^\dagger=\sqrt3(Z_1+Z_2)$.

Squaring $H_0$ using Pauli multiplication rules gives the following
identity for the three-spin triangle:
\begin{equation}
K=3\Id-\frac12H_0^2.
\label{eq:joint-diagonal}
\end{equation}
Thus $K$ commutes with $H_0$ and is diagonalized by the same
transformation $W_0$. Substituting the diagonal form of $H_0$ gives
\begin{equation}
W_0KW_0^\dagger
=3\Id-\frac32(Z_1+Z_2)^2
=-3Z_1Z_2.
\label{eq:exchange-diagonal}
\end{equation}
In this basis, the full Hamiltonian~\eqref{eq:uniform-model}
is a sum of three commuting terms:
\begin{equation}
W_0HW_0^\dagger=\sqrt3D(Z_1+Z_2)-3JZ_1Z_2.
\label{eq:uniform-diagonal}
\end{equation}
A single $ZZ$ rotation is added to the two local $R_z$ rotations,
giving the exact evolution
\begin{equation}
e^{-\ii tH}=W_0^\dagger
R_z^{(1)}(2\sqrt3Dt)R_z^{(2)}(2\sqrt3Dt)
R_{ZZ}^{(12)}(-6Jt)W_0,
\label{eq:encoder-full}
\end{equation}
where $R_{ZZ}^{(ij)}(\theta)=e^{-\ii\theta Z_iZ_j/2}$.

\begin{figure}[!ht]
\centering
\includegraphics[width=.96\linewidth]{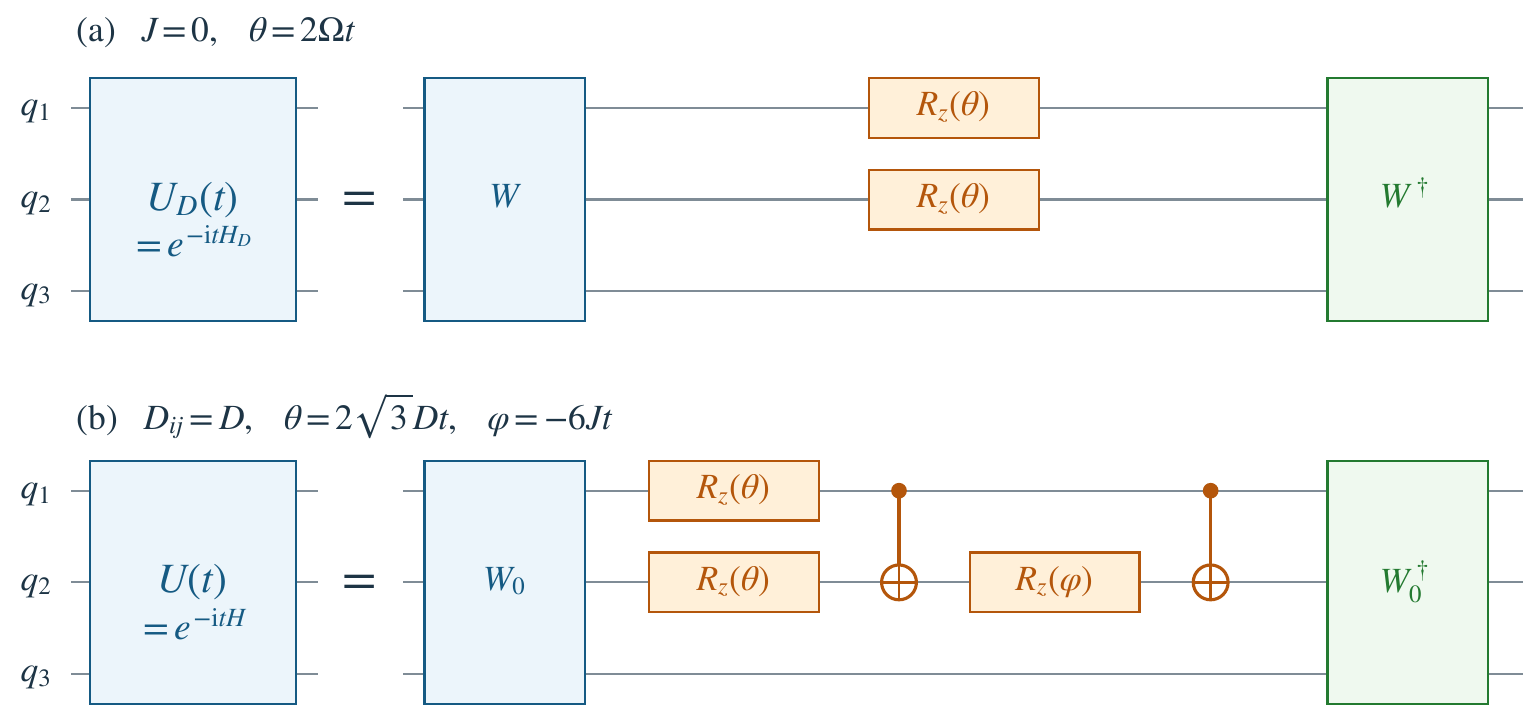}
\caption{Exact evolution under the Hamiltonian~\eqref{eq:Hfull}
using the basis change shown in Fig.~\ref{fig:encoder-four}.
(a) Pure DM, $J=0$, with arbitrary couplings: 8~CNOT gates.
(b) Equal Heisenberg--DM bonds, with $W_0=W(1,1,1)$: 10~CNOT gates.
The central rotations describe evolution in the transformed basis;
the CNOT pair in (b) implements $R_{ZZ}^{(12)}(\varphi)$.}
\label{fig:diagonal-circuits}
\end{figure}

Figure~\ref{fig:diagonal-circuits}(b) shows this circuit with the
central $ZZ$ rotation expanded into elementary gates.
The two basis changes cost 4~CNOT gates each,
and the $ZZ$ rotation adds 2, giving
at most 10~CNOT gates for any real $J,D,t$.
The transformation $W_0$ is fixed: $J,D,t$ enter only the angles of
the three central rotations.

\FloatBarrier

\subsection{Unequal DM couplings}
\label{sec:unequal}

An exact circuit can also be constructed for the Hamiltonian~\eqref{eq:Hfull}
with unequal DM couplings and the same nonzero exchange $J$ on all three
bonds, provided that
\begin{equation}
J^2=D_{12}D_{23}+D_{23}D_{31}+D_{31}D_{12}.
\label{eq:fluxsurface}
\end{equation}
Under this condition, there exist local rotations about the $z$ axis,
$L=\prod_{i=1}^3 R_z^{(i)}(\phi_i)$, that eliminate the symmetric
transverse exchange $X_iX_j+Y_iY_j$ on every edge.
The transverse part of each bond then contains only the antisymmetric
interaction $\widetilde D_{ij}(X_iY_j-Y_iX_j)$ and is therefore purely DM.
The longitudinal terms $JZ_iZ_j$ are unchanged by these rotations.
The transformed Hamiltonian thus takes the form
\begin{equation}
\begin{aligned}
LHL^\dagger&=\widetilde H_D+JQ,\\
\widetilde H_D&=\sum_{(ij)\in\{12,23,31\}}\widetilde D_{ij}(X_iY_j-Y_iX_j),\\
Q&=Z_1Z_2+Z_2Z_3+Z_3Z_1.
\end{aligned}
\label{eq:gauge-reduced}
\end{equation}

To see why this transformation is possible precisely under
condition~\eqref{eq:fluxsurface}, associate the transverse interaction
on an oriented edge $i\to j$ with the complex coefficient
$z_{ij}=J+\ii D_{ij}$. Its real part specifies symmetric exchange,
and its imaginary part specifies the DM interaction. Retaining only
the DM term requires making this coefficient purely imaginary.

Local rotations transform the coefficients as
$z_{ij}\mapsto z_{ij}e^{\ii(\phi_j-\phi_i)}$.
The added angle differences cancel around the triangle, leaving both
the product of the coefficients and its phase
$\Phi=\arg(z_{12}z_{23}z_{31})$ unchanged.
Two independent angle differences allow two bond coefficients to be
made purely imaginary. The third coefficient is then also purely
imaginary if the original product has phase $\Phi=\pi/2\pmod\pi$.
For $J\ne0$, this requirement is equivalent to
condition~\eqref{eq:fluxsurface}.

Explicit expressions for the angles $\phi_i$ and modified couplings
$\widetilde D_{ij}$ are given in Eqs.~\eqref{eq:gaugephases}
and~\eqref{eq:gaugeDM} of Appendix~\ref{app:circuits}.

The operator $Q$ is constant within each sector of fixed excitation
number, and the DM interaction preserves this number.
Hence $[Q,\widetilde H_D]=0$ for arbitrary $\widetilde D_{ij}$.
The exponential of the sum in
Eq.~\eqref{eq:gauge-reduced} therefore factors exactly. Returning
to the original basis gives
\begin{equation}
e^{-\ii tH}=L^\dagger e^{-\ii t\widetilde H_D}e^{-\ii tJQ}L.
\label{eq:gauge-evolution}
\end{equation}
For the DM factor, we use the circuit~\eqref{eq:encoder-dm} already
constructed, with the basis change
$W(\widetilde D_{12},\widetilde D_{23},\widetilde D_{31})$.
It requires at most 8~CNOT gates, as in Fig.~\ref{fig:diagonal-circuits}(a).
The factor $e^{-\ii tJQ}$ is implemented by three commuting rotations
$R_{ZZ}^{(ij)}(2Jt)$ at 2~CNOT gates each, adding 6~CNOT gates.
The rotations $L$ and $L^\dagger$ are single-qubit gates, so the full
circuit requires at most 14~CNOT gates.
This is an upper bound for this construction. At the intersection with
equal bonds, $D_{ij}=D$ and $J^2=3D^2$, the circuit in
Sec.~\ref{sec:uniform} requires only 10~CNOT gates.

The representation $e^{-\ii tH}=W^\dagger e^{-\ii tH_{\rm diag}}W$
uses the same time-independent conjugation principle as Cartan-based
Hamiltonian simulation \cite{Kokcu2022Cartan}.
Here we give an explicit synthesis for the DM triangle: a four-CNOT
basis change, analytical angles valid also at zero couplings, and the
simple normal form $H_{\rm diag}=\Omega(Z_1+Z_2)$.
The identity $K=3\Id-H_0^2/2$ and a local phase condition then yield
the two extensions with uniform Heisenberg exchange.
These results specify short circuits for the stated interaction families.
The CNOT counts given in this section for unrestricted pair connectivity
are upper bounds and are not claimed to be optimal.

\FloatBarrier
\section{Exact decomposition into five DM gates}
\label{sec:five}

Tunable two-spin DM gates provide an alternative to the CNOT circuits
above when they are available on all three logical bonds of the triangle.
Denote an individual DM gate by
\begin{equation}
\Ugate{ij}{\theta}=e^{-\ii\theta(X_iY_j-Y_iX_j)},
\label{eq:Dgate}
\end{equation}
where $\theta$ is a dimensionless gate angle.
For an isolated bond evolving for time $t$, this angle is $\theta=tD_{ij}$.

\begin{figure}[!ht]
\centering
\includegraphics[width=.99\linewidth]{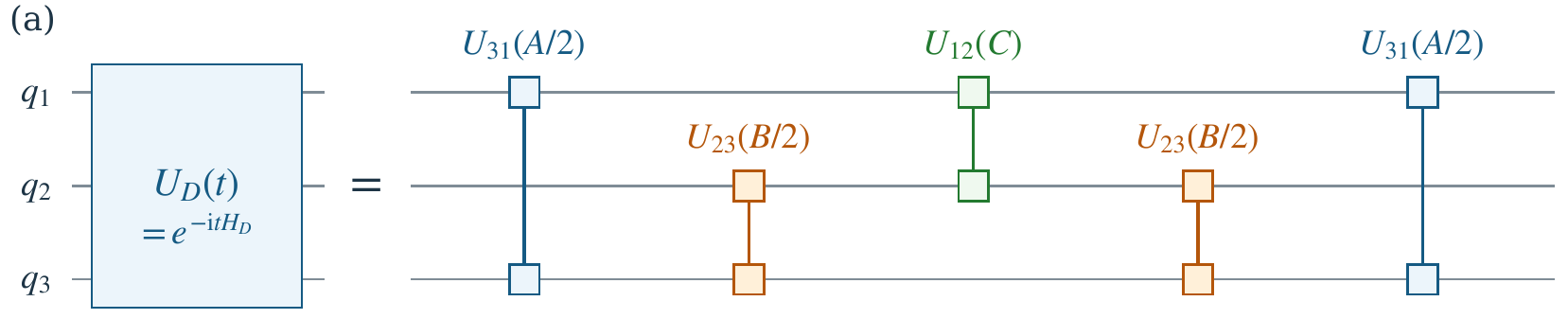}\\[3mm]
\includegraphics[width=.99\linewidth]{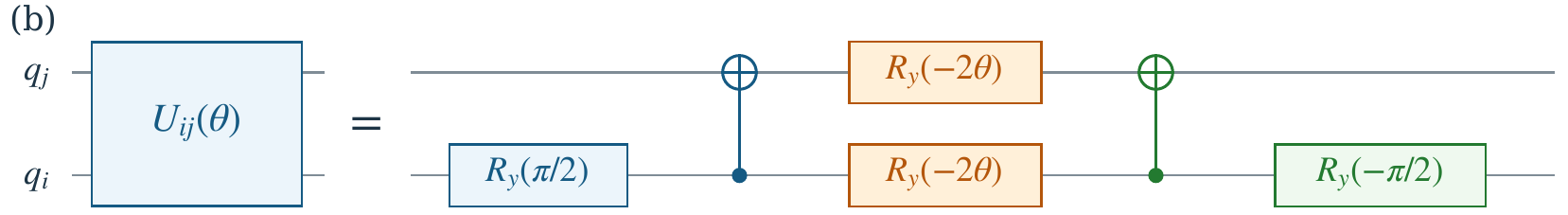}
\caption{(a) Exact evolution of the whole DM triangle for the analytical
angles $A,B,C$ of Eq.~\eqref{eq:angles-general}.
(b) Implementation of a single DM gate with 2~CNOT gates.}
\label{fig:circuits}
\end{figure}

One symmetric Strang step over the three bonds consists of five such gates.
For the sequence in Fig.~\ref{fig:circuits}, it corresponds to the
linear parameters
\[
A_{\mathrm S}=tD_{31},\qquad B_{\mathrm S}=tD_{23},\qquad C_{\mathrm S}=tD_{12}.
\]
This is a second-order approximation: the operator error
of one step is $O(t^3)$ as $t\to0$ at fixed couplings.

For the DM triangle, this sequence becomes exact when the linear
parameters are replaced by the analytical angles $A,B,C$ derived in
Appendix~\ref{app:circuits}, Eq.~\eqref{eq:angles-general}.
For arbitrary real $D_{ij}$ and $t$, we obtain
\begin{equation}
e^{-\ii tH_D}=
\Ugate{31}{A/2}\Ugate{23}{B/2}\Ugate{12}{C}
\Ugate{23}{B/2}\Ugate{31}{A/2}.
\label{eq:identity}
\end{equation}
The gate count and order remain the same; only the angles change.
The identity holds on the full Hilbert space.
When implemented through CNOTs, the five DM blocks require 10~CNOT gates;
this form is particularly useful with tunable native exchange operations
\cite{BaiMarvian2024}.

A palindromic sequence of the same bond structure was used by Lu et al.\
\cite{Lu2016Chiral}; merging its two central gates gives five bond operations.
Here we give closed-form angles for arbitrary real DM couplings and time,
with an identity on the full three-spin Hilbert space.

\FloatBarrier
\Needspace{7\baselineskip}
\section{Two physical examples}
\label{sec:examples}

\begin{figure}[!ht]
\centering
\includegraphics[width=\linewidth]{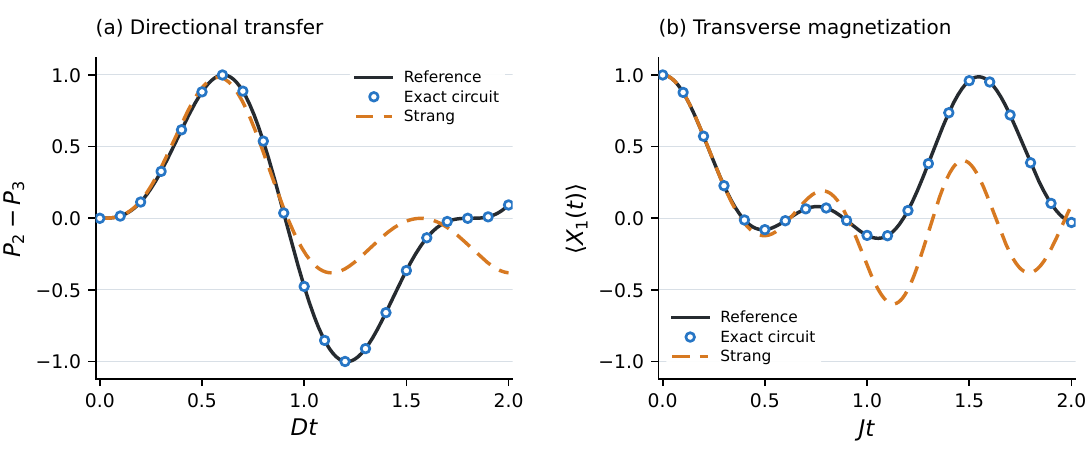}
\caption{(a) Directional transfer in a pure DM triangle: $J=0$, all
$D_{ij}=D>0$, initial state $|100\rangle$.
(b) Transverse magnetization for the parameters in Eq.~\eqref{eq:example}
from $|{+}00\rangle$.
Solid lines: exact reference evolution. Open markers: exact circuits
based on a change of basis. Dashed lines: one symmetric Strang step
for the three bond Hamiltonians. The dashed curve in panel (a) uses the linear parameters
$A_{\mathrm S},B_{\mathrm S},C_{\mathrm S}$. Replacing them by the analytical
angles in Eq.~\eqref{eq:angles-general} reproduces the exact curve
without changing the circuit structure.
The exact circuits require at most 8 and
14~CNOT gates in (a) and (b); the Strang circuits require at most
10 and 15, respectively.}
\label{fig:dynamics}
\end{figure}

To demonstrate the exact circuits, we consider two examples of spin-triangle
dynamics (Fig.~\ref{fig:dynamics}). In the first, the DM interaction
produces directional transfer of an initially localized excitation
between spins. In the second, we follow transverse magnetization in the
model with both Heisenberg exchange and DM interactions: this observable is sensitive to relative phases
between states with different excitation numbers.

\Needspace{7\baselineskip}
\paragraph{Directional transfer.}
Set $J=0$, $D_{12}=D_{23}=D_{31}=D>0$, and start from $|100\rangle$.
The probability of an excitation on spin $i$ is
$P_i(t)=\langle(\Id-Z_i)/2\rangle$.
To observe directional transfer, we compare the populations of the
other two spins through $P_2-P_3$. Direct exponentiation of $H_D$ in the one-excitation sector gives
\begin{equation}
P_2(t)-P_3(t)=\frac{4}{3\sqrt3}\sin(2\sqrt3Dt)
\bigl[1-\cos(2\sqrt3Dt)\bigr].
\label{eq:transfer-signal}
\end{equation}
At $Dt=\pi/(3\sqrt3)$ the excitation is entirely on spin 2;
at $Dt=2\pi/(3\sqrt3)$ it is entirely on spin 3.
At $Dt=\pi/\sqrt3$, the system returns to its initial state. Reversing the sign
of all DM couplings reverses this sequence.

\paragraph{Transverse magnetization.}
Take an example from the family~\eqref{eq:fluxsurface}:
\begin{equation}
J=1,\qquad (D_{12},D_{23},D_{31})=(1/2,1,1/3).
\label{eq:example}
\end{equation}
The initial state
$|{+}00\rangle=(|000\rangle+|100\rangle)/\sqrt2$
is a coherent superposition of the zero- and one-excitation sectors.
The Hamiltonian does not mix these sectors, but their components
accumulate a relative phase. The term $JQ$ in
Eq.~\eqref{eq:gauge-reduced} has eigenvalue $3J$ on $|000\rangle$
and $-J$ in the one-excitation sector, contributing $4Jt$ to this phase.
For this state, $\langle X_1(t)\rangle$ is sensitive to the phase
between the amplitudes of $|000\rangle$ and $|100\rangle$,
which populations alone do not determine.
For the parameters in Eq.~\eqref{eq:example}, the exact signal is
\begin{equation}
\langle X_1(t)\rangle=
\frac{72+85\cos(\sqrt{157}\,t/3)}{157}\cos(4t).
\label{eq:magnetization-signal}
\end{equation}
For $\langle Y_1(t)\rangle$, the factor $\cos(4t)$ is replaced by
$\sin(4t)$. Both components are derived in Sec.~S1 of the Supplemental
Material and measured in Sec.~\ref{sec:hardware}.

The solid curves in Fig.~\ref{fig:dynamics} are obtained by direct
evaluation of $e^{-\ii tH}$, whereas the markers are obtained by
multiplying the single-qubit rotation and CNOT matrices in circuit order.
Both calculations use the full eight-dimensional state space,
without hardware noise or statistical sampling of measurement outcomes,
and agree with the analytical expressions in
Eqs.~\eqref{eq:transfer-signal} and~\eqref{eq:magnetization-signal}.

The reference Strang step contains five bond exponentials;
each costs at most 2~CNOT gates for pure DM and 3 for the full model
\cite{VatanWilliams2004TwoQubit,Shende2004Minimal}.

\FloatBarrier

\FloatBarrier
\section{Exact triangle evolution on a three-qubit chain}
\label{sec:path}

The constructions above assume that two-qubit gates are available on
all three pairs. On superconducting processors with restricted connectivity,
such as IBM heavy-hex devices \cite{IBMHeavyHex2021}, three connected
qubits can form a chain with no direct connection between its ends.
We therefore seek a basis change using only neighboring pairs in a
chain $1$--$2$--$3$. The simulated Hamiltonian retains all three
interactions; only the allowed gate connections change.

\begin{figure}[!ht]
\centering
\includegraphics[width=\linewidth]{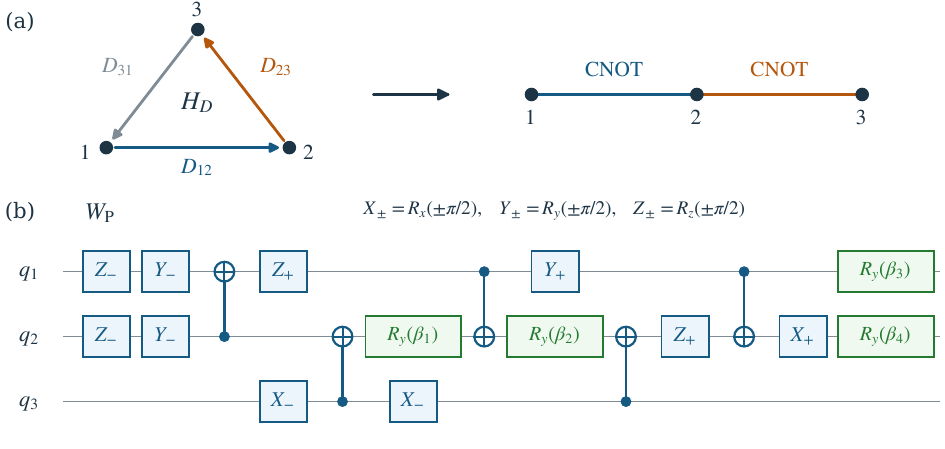}
\caption{Exact triangle diagonalization on a three-qubit chain.
(a) The logical Hamiltonian retains all three oriented DM bonds;
two-qubit gates use only pairs $12$ and $23$.
(b) The basis change $W_{\rm P}$ contains five CNOT gates. Fixed quarter-turns are denoted
$X_\pm=R_x(\pm\pi/2)$, and likewise for $Y_\pm,Z_\pm$.
Blue gates are fixed; green rotations use Eq.~\eqref{eq:path-angles}.}
\label{fig:path-circuit}
\end{figure}

Figure~\ref{fig:path-circuit} gives an explicit solution
$W_{\rm P}(D_{12},D_{23},D_{31})$.
With $a=D_{12}$, $b=D_{23}$, $c=D_{31}$,
$r_{ac}=\sqrt{a^2+c^2}$, and $r_{ab}=\sqrt{a^2+b^2}$,
its four adjustable angles are
\begin{equation}
\begin{aligned}
\beta_1&=\pi-\atanTwo(c,a),&
\beta_2&=-\atanTwo(b,a),\\
\beta_3&=\pi+\atanTwo(b,r_{ac}),&
\beta_4&=-\pi-\atanTwo(c,r_{ab}).
\end{aligned}
\label{eq:path-angles}
\end{equation}
With $\atanTwo(0,0)=0$, this circuit satisfies
\begin{equation}
W_{\rm P}H_DW_{\rm P}^\dagger=\Omega(Z_1+Z_2),\qquad
\Omega=\sqrt{a^2+b^2+c^2},
\label{eq:path-diagonal}
\end{equation}
for all real couplings, including zeros.
A short derivation is given in Appendix~\ref{app:path}.

Exact DM evolution follows by inserting two $z$ rotations between
$W_{\rm P}$ and $W_{\rm P}^\dagger$,
\begin{equation}
e^{-\ii tH_D}=W_{\rm P}^\dagger
R_z^{(1)}(2\Omega t)R_z^{(2)}(2\Omega t)W_{\rm P}.
\label{eq:path-pure}
\end{equation}
Thus restricting the connectivity adds only one CNOT to the basis
change, or two to the full evolution, without SWAP gates.
The couplings fix the basis change; only the two central angles depend
on time. For equal nonzero couplings, this basis change uses the minimum
number of CNOT gates for the target diagonal form in Eq.~\eqref{eq:path-diagonal}
(Supplemental Material, Sec.~S4).

The two families with Heisenberg exchange also admit circuits on the
chain. Their costs are summarized in Table~\ref{tab:connectivity-cost},
with explicit circuits in Appendix~\ref{app:path-recipes}.
The next section uses the chain circuits for pure DM and the
phase-constrained family in experiments on an IBM quantum processor.

\begin{table}[!ht]
\centering\small
\caption{Constructive CNOT upper bounds for exact three-spin evolution.
Initial-state preparation and measurement are excluded. Special
parameters or times may require fewer gates.}
\label{tab:connectivity-cost}
\begin{tabular}{@{}lcc@{}}
\hline
Interaction family & Any pair & Chain\\
\hline
Pure DM, arbitrary $D_{ij}$ & 8 & 10\\
Uniform $J,D$ & 10 & $11^{\rm a}$\\
Uniform $J\ne0$, Eq.~\eqref{eq:fluxsurface} & 14 & 16\\
\hline
\end{tabular}

\smallskip
\begin{minipage}{0.9\linewidth}\footnotesize
$^{\rm a}$Includes local recompilation of a two-qubit fragment
(Supplemental Material, Sec.~S3.2).
\end{minipage}
\end{table}

\FloatBarrier

\Needspace{10\baselineskip}
\section{Spin-triangle dynamics on a quantum processor}
\label{sec:hardware}

We tested the analytically exact circuits on the 156-qubit superconducting
quantum processor IBM Kingston (\texttt{ibm\_kingston}) \cite{IBMQuantum2026}
by measuring transverse-magnetization dynamics and directional state transfer.
The circuits implement all three triangle interactions
on a three-qubit chain without SWAP gates. We compare them with direct
bondwise splitting into $r$ symmetric second-order Trotter steps
(the Strang formula), using the same initial
states, model parameters, physical placements, and measurement settings.
The native two-qubit gate is CZ, related to CNOT by single-qubit basis
changes; circuit costs below are therefore expressed as CZ counts.
Each plotted per-chain estimate uses 2048 shots per measurement basis,
without readout mitigation, postselection, or phase fitting.
The protocol, simplifications of the routed Strang circuits, population
dynamics, and results for all tested chains
are given in the Supplemental Material, Sec.~S5.

For exchange and unequal DM couplings, we choose $J=1$,
$(D_{12},D_{23},D_{31})=(1/2,1,1/3)$ and the initial state
$|{+}00\rangle$. Measurements of $\langle X_1(t)\rangle$ and
$\langle Y_1(t)\rangle$ probe the relative phase of the zero- and
one-excitation branches. To distinguish changes in the length and
direction of the transverse magnetization, we write
\[
C_1(t)=\langle X_1(t)\rangle+\ii\langle Y_1(t)\rangle,
\qquad \varphi_1(t)=\arg C_1(t).
\]
The modulus changes even in ideal evolution as the excitation moves
between spins; its decrease alone does not indicate decoherence.
Ideally, $C_1(t)=a_1(t)e^{4\ii t}$, where $a_1(t)$ is the real prefactor
of $\cos(4t)$ in Eq.~\eqref{eq:magnetization-signal}.
At $Jt\simeq0.618$ and $0.886$, this amplitude changes
sign through zero. The ideal phase changes by $\pi$ and is undefined
at the zero itself, while the ideal components
$\langle X_1(t)\rangle$ and $\langle Y_1(t)\rangle$ remain continuous.
Analytical expressions are given in the Supplemental Material, Sec.~S1.
Figure~\ref{fig:hardware-dynamics} shows the modulus
and phase on one chain in two separate experiments: Strang at $r=8$
and $r=16$, each with its own exact-circuit control. This fixed chain
serves as an illustration; its selection criterion and results for the
other chains are given in the Supplemental Material. The exact circuit
uses 16 CZ gates, compared with 105 and 201 for Strang at $r=8$ and
$r=16$, respectively. Differences between the two experiments cannot be
attributed solely to the step count.

\begin{figure}[!ht]
\centering
\includegraphics[width=\linewidth]{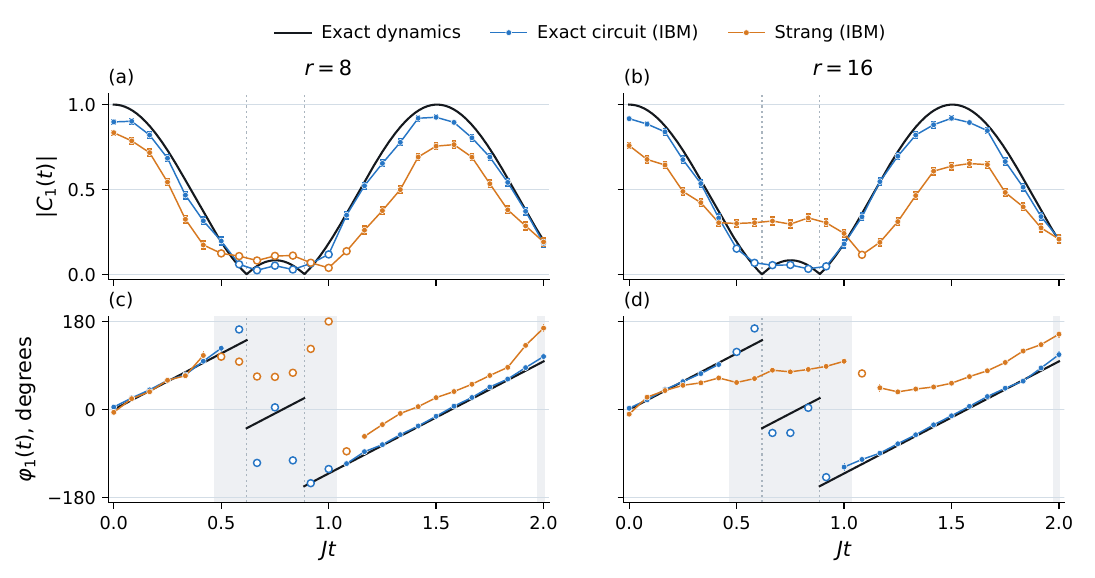}
\caption{Amplitude and phase of the transverse magnetization on the IBM
quantum processor for $J=1$, $(D_{12},D_{23},D_{31})=(1/2,1,1/3)$, from
$|{+}00\rangle$. Top: $|C_1(t)|$; bottom: $\varphi_1(t)$.
(a,c) Comparison with Strang at $r=8$; (b,d) comparison with Strang at $r=16$.
Black lines show exact dynamics; blue and orange points denote the
exact circuit and Strang, respectively. Vertical dashed lines mark zeros
of the ideal signal; shading denotes $|C_{1,\mathrm{ideal}}|<0.25$.
Open markers flag a measured signal-to-noise ratio below 5.
Error bars show one statistical standard error; processing details
are given in the Supplemental Material, Sec.~S5.}
\label{fig:hardware-dynamics}
\end{figure}

\Needspace{4\baselineskip}
The exact circuit reproduces the amplitude revival and subsequent
phase evolution. For Strang, substantial phase deviations persist
after a strong signal has recovered. Here the phase error is the
magnitude of the shortest angular deviation from exact dynamics.
For example, at the single sampled time $Jt=4/3$ on the displayed chain,
it is approximately $46^\circ$ for $r=8$ and $96^\circ$ for $r=16$,
compared with about $3^\circ$ for the exact circuit in both experiments. At the sampled times in $1.25\le Jt\le1.75$, the phase
error of the ideal Strang formula remains below $3.1^\circ$.
The observed discrepancy therefore extends beyond a simple loss of
contrast: circuit execution also distorts the direction of the
transverse magnetization. These measurements do not isolate a specific
error mechanism.

\Needspace{6\baselineskip}
The experiment providing the $r=16$ data also included Strang with
$r=4$. Over the full measured time series, the shorter circuit had a
smaller root-mean-square error in the complex signal $C_1(t)$.
The exact circuit had a smaller error than either Strang variant on
every chain. Results for all chains and an independent repeat are given
in the Supplemental Material, Sec.~S5.

\FloatBarrier
Directional transport also enables transfer of a single-qubit state.
For $J=0$, $D_{12}=D_{23}=D_{31}=\pm1$, and $t_*=\pi/(3\sqrt3)$, denote the two
target propagators by $U_\pm$. For any
$|\psi\rangle=\alpha|0\rangle+\beta|1\rangle$, ideal evolution gives
\[
U_+(t_*)|\psi00\rangle=|0\psi0\rangle,\qquad
U_-(t_*)|\psi00\rangle=|00\psi\rangle.
\]
No additional relative-phase correction is required. For each method
and sign of DM, the same evolution circuit is used for all input states.
The exact circuit uses 10 CZ gates, compared with 88 for Strang at $r=8$.
In the experiment, spin 1 was prepared in the six Pauli eigenstates
$|{+X}\rangle$, $|{-X}\rangle$, $|{+Y}\rangle$, $|{-Y}\rangle$,
$|{+Z}\rangle$, and $|{-Z}\rangle$, with the other two spins initially
in $|00\rangle$. The three measurement settings $XXX$, $YYY$, and $ZZZ$
provide estimates of all three Bloch-vector components of every output spin
and hence of its single-qubit density matrix $\rho_i$ (local tomography).
The transfer fidelity $F_{\psi,i}=\langle\psi|\rho_i|\psi\rangle$ is
evaluated directly from these components; for example, input $|\pm X\rangle$
gives $(1\pm\langle X_i\rangle)/2$.

\begin{figure}[!ht]
\centering
\includegraphics[width=0.88\linewidth]{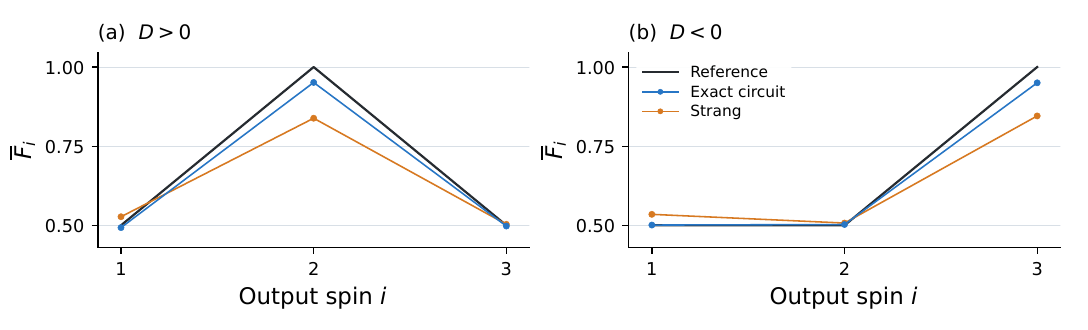}
\caption{Directional state transfer on the IBM quantum processor at $J=0$
and $t_*=\pi/(3\sqrt3)$.
(a) $D_{ij}=+1$, with spin 2 as receiver;
(b) $D_{ij}=-1$, with spin 3 as receiver.
$\overline F_i$ averages $\langle\psi|\rho_i|\psi\rangle$ over the six
inputs $|\psi\rangle$ ($\pm X,\pm Y,\pm Z$) and five chains,
where $\rho_i$ is the state of output spin $i$.
Black lines show ideal evolution; blue and orange points denote the
exact circuit and Strang with $r=8$, respectively.
Error bars show one statistical standard error.}
\label{fig:hardware-state-transfer}
\end{figure}

\Needspace{4\baselineskip}
Figure~\ref{fig:hardware-state-transfer} shows that
changing the sign of DM switches the receiver. Both circuits reproduce
directional transfer, with higher fidelity for the exact circuit:
averaged over the six inputs and five chains, the transfer fidelities
are $95.1\%$ for $1\to2$ and $95.0\%$ for $1\to3$, compared with
$83.8\%$ and $84.6\%$ for Strang. The advantage also holds for the
four equatorial inputs $\pm X,\pm Y$: they have equal $|0\rangle$ and
$|1\rangle$ populations but different relative phases, and thus test
the transfer of phase information. The reported fidelities include state-preparation and
measurement errors and describe single-qubit state transfer rather
than the full three-qubit operator. Ideal mean Strang transfer
fidelity exceeds $0.99999$ in both directions: the observed difference
arises when the circuits are executed on the processor.

\FloatBarrier

\section{Exact triangle blocks in kagome simulation}
\label{sec:kagome}

We now test numerically whether the resource advantage of exact triangle
circuits persists in a larger system whose neighboring blocks no longer
commute. This calculation assumes ideal gates and unrestricted pair
connectivity. The kagome lattice, a two-dimensional network of
corner-sharing triangles,
provides a natural test. This geometry is used in models of frustrated
magnetism, and its Hamiltonian can be assembled from the three-spin
blocks considered above.

Shared spins between neighboring triangles leave a splitting error even
when evolution within each block is exact. On a small cluster, we compare
the cost of triangle and bond decompositions at the same target accuracy
for the full evolution.

Divide the triangles into two families according to their orientation,
upward ($\triup$) and downward ($\tridown$):
\begin{equation}
H=A+\varepsilon B,\qquad
A=\sum_{\triup}h_{\triup}(J,D),\quad
B=\sum_{\tridown}h_{\tridown}(J,D),
\label{eq:kagome-model}
\end{equation}
Here each $h$ is the Hamiltonian~\eqref{eq:Hfull} with equal couplings on
one triangle. The parameter $\varepsilon$ sets the relative coupling
strength: upward triangles have couplings $(J,D)$ and downward
triangles $(\varepsilon J,\varepsilon D)$. Triangles in the same family
share no spins, so their Hamiltonians commute. The evolution generated
by either family therefore factors exactly into the triangle circuits
of Sec.~\ref{sec:uniform}. Grouping interactions into exact triangle blocks has been
used previously \cite{NegishiYang2026,Maskara2023Programmable}.

For total time $T$ and $r$ steps, let $\delta=T/r$.
We first illustrate how the two families alternate using a second-order
formula; the numerical comparison below uses the same fourth-order
formula for both decompositions. The Strang step is
\begin{equation}
S_2(\delta)=e^{-\ii\delta A/2}
 e^{-\ii\varepsilon\delta B}e^{-\ii\delta A/2},\qquad
U(T)\simeq S_2(\delta)^r.
\label{eq:kagome-strang}
\end{equation}
In this construction, the error arises from the noncommutation of $A$
and $B$ \cite{Childs2021Trotter}, while evolution within each triangle
is implemented exactly. At $\varepsilon=0$ only independent triangles
remain and the splitting becomes exact.

\begin{figure}[t]
\centering
\includegraphics[width=.92\linewidth]{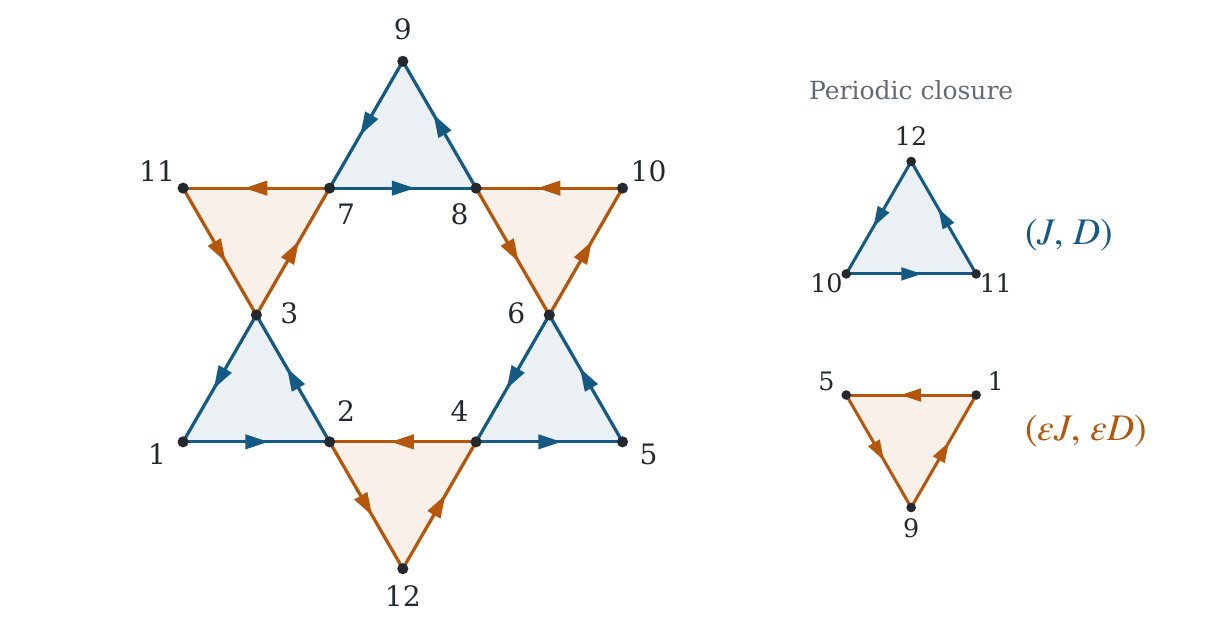}
\caption{The 12-spin periodic kagome cluster used in the calculation.
The star shows six triangles; the two triangles on the right complete
the periodic bonds. Repeated node labels denote the same spins,
so the full cluster contains four upward and four downward triangles.
Blue and orange triangles carry couplings $(J,D)$ and
$(\varepsilon J,\varepsilon D)$, respectively.
Arrows specify the cyclic DM orientations.}
\label{fig:kagome}
\end{figure}

For a numerical illustration, consider the periodic 12-spin cluster
with 24 bonds shown in Fig.~\ref{fig:kagome}. Set $J=1$ and $D=0.2$. We take two regimes: weaker downward couplings at
$(\varepsilon,T)=(0.2,5)$ and a uniform lattice at
$(\varepsilon,T)=(1,1)$. Both satisfy $\varepsilon JT=1$: the evolution time is increased as the
downward couplings weaken, so that dynamics on their timescale can develop.

For the numerical comparison, we use the symmetric
fourth-order Omelyan formula \cite{Bosse2025THRIFT}. This tests the benefit
of exact triangle blocks within a higher-order splitting between blocks.
The same formula is applied either to the two triangle families or to
four fixed layers of disjoint bonds.
A triangle block uses 10~CNOT gates and a bond exponential uses 3.
A layer of four triangles therefore costs 40~CNOT gates,
and a layer of six bonds costs 18.
Routing on a specific processor can add gates.
After neighboring equal layers are merged, $r$ steps require
$40(8r+1)$ CNOT gates for the triangle decomposition and
$18(24r+1)$ CNOT gates for the bond decomposition.
Using two triangle layers instead of four bond layers lowers the cost
of the repeated step from 432 to 320 CNOT gates. A more expensive local
exponential can therefore make the full sequence cheaper; a reduction
in the required step count provides a further saving.

Let $U(T)=e^{-\ii TH}$ denote the exact evolution operator of the full
Hamiltonian, and let $C$ denote the unitary operator of the approximate
circuit with a chosen decomposition and number of steps.
Define the maximum state error by
\begin{equation}
E_{\max}(C)=\max_{\|\psi\|=1}
\left[1-|\langle\psi|U(T)^\dagger C|\psi\rangle|^2\right].
\label{eq:kagome-emax}
\end{equation}
The expression inside the modulus is the overlap between the exact
state $U(T)|\psi\rangle$ and the circuit output $C|\psi\rangle$.
The maximum is taken over the full Hilbert space, including
superpositions of states from different excitation sectors.
Table~\ref{tab:kagome-cost} gives examples of circuits reaching
$E_{\max}\le10^{-2}$. The explicit Omelyan formula and its coefficients, layer sequences,
error calculation and additional thresholds are provided in Sec.~S2
of the Supplemental Material and in the code package.

\begin{table}[H]
\centering
\caption{Illustrative circuits reaching $E_{\max}\le10^{-2}$.
CNOT counts are rounded to two significant figures.}
\label{tab:kagome-cost}
\small
\begin{tabular}{@{}crrrr@{}}
\hline
$(\varepsilon,T)$ & \multicolumn{2}{c}{Steps $r$} &
\multicolumn{2}{c}{CNOT count} \\
 & Triangles & Bonds & Triangles & Bonds \\
\hline
$(0.2,5)$ & 13 & 17 & $4.2\times10^3$ & $7.4\times10^3$ \\
$(1,1)$ & 5 & 5 & $1.6\times10^3$ & $2.2\times10^3$ \\
\hline
\end{tabular}
\end{table}

In the regimes considered, exact triangle blocks reduce the CNOT count
by approximately 43\% and 25\%, respectively. With weaker downward couplings they also reduce
the number of steps; in the uniform case the step count is the same,
but each step is cheaper. The total cost is still several thousand
CNOT gates. These examples show that the exact circuits can be useful
as local blocks in simulations of larger systems; the gain depends on
the model, regime and target accuracy.

\FloatBarrier

\section{Coherent-error compensation in triangle blocks}
\label{sec:compensation}

The explicit triangle circuit also provides a compact parametrization
for compensating control errors. We consider a constant unwanted
$ZZ$ interaction acting throughout finite-duration gates.
This numerical model includes coherent errors only.
The gate model and optimization procedure are given in the
Supplemental Material, Sec.~S6.

\begin{figure}[!ht]
\centering
\includegraphics[width=.72\linewidth]{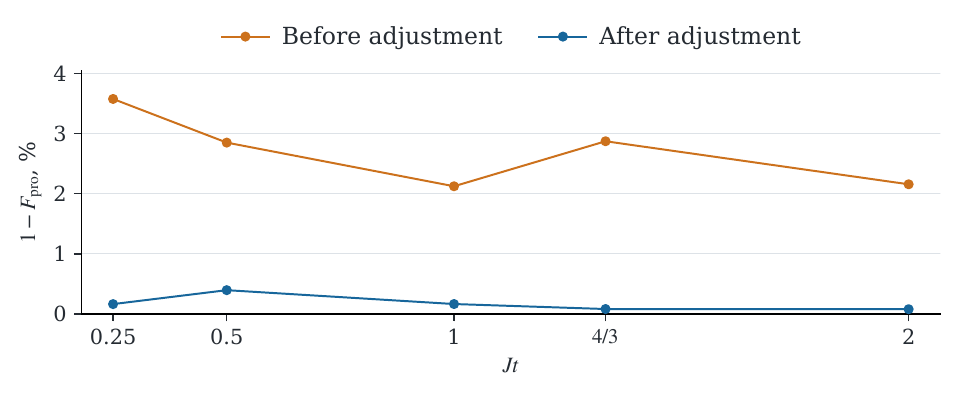}
\caption{Compensating a known $ZZ$ interaction by retuning the existing
$R_z$ angles of one triangle circuit. The simulated evolution time is
$Jt$; the physical duration remains $23.6\tau_{\mathrm{CZ}}$ at every
point. The error is $1-F_{\mathrm{pro}}$, where
$F_{\mathrm{pro}}=|\operatorname{Tr}(U^\dagger V)|^2/8^2$ compares the
target operator $U=e^{-\ii tH_\triangle}$ with the implemented unitary
$V$. Each point uses a separate angle optimization, retaining the
same 16 CZ gates. Lines connect the sampled times.}
\label{fig:single-compensation}
\end{figure}

First consider the exchange--DM triangle with $J=1$ and
$(D_{12},D_{23},D_{31})=(1/2,1,1/3)$, implemented by the 16-CZ chain
circuit. Add an unwanted interaction $K=\kappa Z_1Z_2$, with
$\kappa\tau_{\mathrm{CZ}}=0.03$, where $\tau_{\mathrm{CZ}}$ is the
physical CZ duration. For the numerical estimate, we choose
$\tau_{\mathrm{CZ}}=200$~ns and SX and X durations of
40 and 80~ns. The conditional frequency shift due to ZZ is
$4\kappa/(2\pi)\simeq95.5$~kHz. The gate durations and unwanted
coupling strength are comparable in order of magnitude to published
values for superconducting circuits
\cite{Hashim2022ZZScale,Wei2022ZZScale}.
For each target evolution time, we numerically
retune the existing virtual $R_z$ angles to approximate the full target
operator. This follows the control principle of adjusting a gate
implementation to reduce its coherent error \cite{Khodjasteh2012}.
Across the sampled times in Fig.~\ref{fig:single-compensation}, the
process infidelity falls from $2.12$--$3.57\%$ to $0.080$--$0.397\%$.
Retuning substantially reduces the error of the specified model
without increasing the gate count or physical execution time.

\begin{figure}[!ht]
\centering
\includegraphics[width=\linewidth]{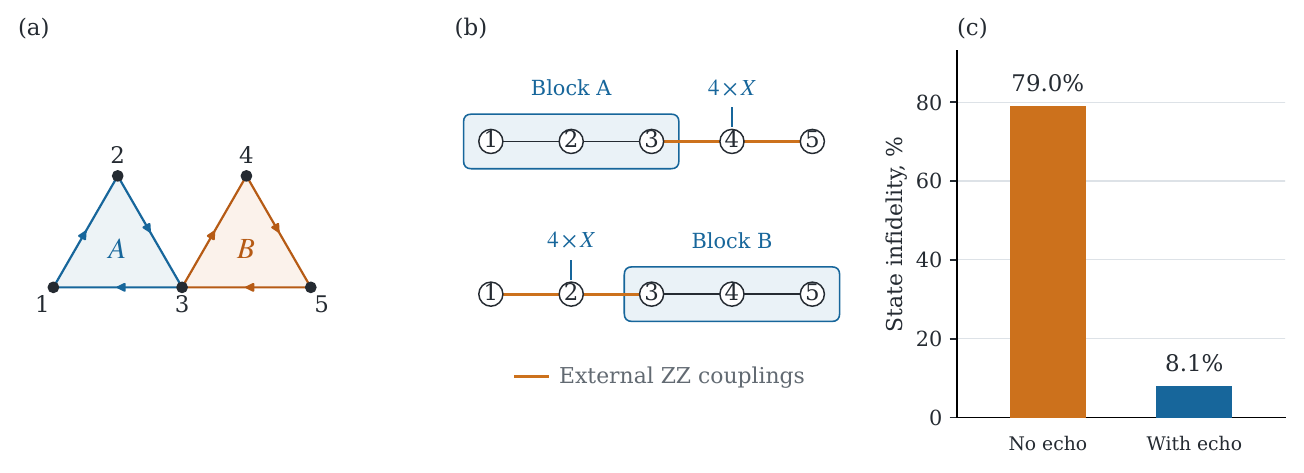}
\caption{Protecting the surroundings of a triangle block in a
five-qubit calculation.
(a) Two triangles sharing spin 3; arrows set the cyclic DM orientations.
(b) Four $X$ pulses act on spin 4 while $A$ is implemented, or on spin 2
while $B$ is implemented. Active-block operations pause during each
finite-duration echo pulse.
(c) Mean infidelity relative to the exact five-spin evolution over
four test inputs, with the same locally calibrated angles in both
cases. It includes the error of splitting between triangles.
Echo increases the physical duration from $118$ to
$126\tau_{\mathrm{CZ}}$ while retaining 80 CZ gates.}
\label{fig:five-compensation}
\end{figure}

When triangle blocks are combined, unwanted interactions with idle
neighbors must also be considered. Take triangles $A=(1,2,3)$ and $B=(3,4,5)$ on a five-qubit
chain, each with the parameters above. The target Hamiltonian is
$H_A+0.2H_B$, and unwanted $ZZ$ interactions of the same strength
act on all four chain bonds.
Two symmetric second-order Trotter steps at $Jt=1$ give five triangle blocks.
Using only angles calibrated for isolated triangles leaves a mean
state infidelity of $79.0\%$ over four test inputs.
Inserting four $X$ echo pulses on an idle qubit during each block,
at gate boundaries selected for this implementation, reduces this
error to $8.1\%$ with the same local angles
(Fig.~\ref{fig:five-compensation}).
The sequence retains 80 CZ gates and adds 20 single-qubit $X$ pulses.
Echo exploits the sign reversal $XZX=-Z$ to suppress unwanted couplings.
This use of pulse timing to suppress unwanted couplings is related to
established control and scheduling methods \cite{Khodjasteh2012,Xie2022}.
The selected echo is specific to the circuit and error model.
The example shows how local compensation and protection of idle
neighbors can work together when exact triangle blocks are composed.

\FloatBarrier

\Needspace{12\baselineskip}
\section{Conclusion}

We have constructed short analytical circuits for the full evolution
of a pure-DM triangle and the two families with Heisenberg exchange
considered here. The circuits are exact at arbitrary evolution time
and for arbitrary initial states, with a gate count independent of time.
They admit implementations on a three-qubit chain without SWAP gates,
retaining all interactions of the simulated triangle.
For pure DM, we also obtain an exact factorization into five pairwise
DM gates.

Experiments on a superconducting IBM quantum processor demonstrate
excitation transfer and reproduce the amplitude and phase of transverse
magnetization. The tested circuits based on the symmetric second-order
Trotter formula exhibit phase deviations at substantial signal amplitude
that cannot be explained by contrast attenuation alone. Short analytically
exact circuits substantially reduce these distortions. In the directional
transfer experiment, the DM sign selects the receiver, and single-qubit
tomography tests the transfer of populations and phase information for
six input states. In both directions, the exact circuit achieves higher
mean transfer fidelity than the tested Trotter circuit.

Circuits for an individual triangle can serve as local operations
in simulations of larger spin systems. In the two regimes studied on a
12-spin kagome cluster, their use reduces the CNOT count under the same
product formula and error threshold. Splitting between neighboring
triangles remains approximate. This comparison assumes unrestricted
pair connectivity and excludes the cost of mapping the full lattice
onto a processor. The solutions expand the set of exactly implementable
local models for digital studies of spin dynamics on frustrated lattices
with DM interactions.

In a separate finite-pulse model, retuning existing angles substantially
reduces a specified coherent ZZ error without increasing the gate count.
In a five-qubit composition, additional echo pulses on inactive neighbors
reduce the error with the same local angles. These examples show how
exact triangle blocks can be combined with compensation of known errors;
transferring such calibration to larger systems requires a separate test.

\Needspace{7\baselineskip}
\section*{Data availability}
Code, numerical and experimental data, and Supplemental Material
are available on Zenodo:
\href{https://doi.org/\DatasetVersionDOI}{doi:\DatasetVersionDOI}.

\section*{Acknowledgments}
The author acknowledges access to IBM Quantum computing resources.
The conclusions are those of the author and do not reflect the official
position of IBM or the IBM Quantum team.

\section*{Manuscript preparation}
ChatGPT was used in preparing the text, translation, and code.

\appendix

\Needspace{13\baselineskip}
\section{Parameters and derivations of the exact circuits}
\label{app:circuits}

The circuit recipes use the angles in Eq.~\eqref{eq:encoder-angles}
for $W$, those in Eq.~\eqref{eq:angles-general} for the five DM gates,
and the local rotations and modified couplings in
Eqs.~\eqref{eq:gaugephases}--\eqref{eq:gaugeDM} for the phase-constrained family.
Sections~\ref{app:path} and~\ref{app:path-recipes} give the chain implementations.
The short derivations below verify these constructions on the full
three-spin Hilbert space.

\subsection{Basis change for the DM triangle}
\label{app:encoder}

The circuit in Fig.~\ref{fig:encoder-four} successively reduces $H_D$
to two independent fields along the $z$ axis.
CNOT gates and fixed local rotations transform the Pauli strings;
two coupling-dependent angles eliminate selected components, and
the other two align the resulting fields.
Numerical search for equal DM couplings suggested the CNOT sequence.
The derivation below determines the angles and proves that this circuit
works for arbitrary real couplings.

Let $G_1,\ldots,G_m$ denote the quantum gates in the figure, ordered
from left to right. Then $W=G_m\cdots G_1$, and each gate acts as
$H\mapsto G_kHG_k^\dagger$.
Set $a=D_{12}$, $b=D_{23}$, $c=D_{31}$,
$r_{ab}=\sqrt{a^2+b^2}$, $r_{bc}=\sqrt{b^2+c^2}$ and
$\Omega=\sqrt{a^2+b^2+c^2}$.

After the first three CNOT gates and their preceding single-qubit rotations,
the coefficients of $Y_2$ and $Y_1Y_3$ vanish if
\[
a\sin\alpha_1-b\cos\alpha_1=0,\qquad
b\cos\alpha_2+c\sin\alpha_2=0.
\]
With the angle branches specified below, the remaining combinations are
$a\cos\alpha_1+b\sin\alpha_1=r_{ab}$ and
$b\sin\alpha_2-c\cos\alpha_2=r_{bc}$. This gives
\begin{equation}
H_3=cX_2Y_3-r_{ab}Z_2+r_{bc}X_1+aZ_1.
\label{eq:encoder-intermediate}
\end{equation}
Next, $R_z^{(3)}(\pi/2)$ followed by $\mathrm{CX}_{2\to3}$ replaces
$cX_2Y_3$ by $-cX_2$, giving two independent fields:
\[
(r_{bc}X_1+aZ_1)+(-cX_2-r_{ab}Z_2).
\]
These two gates commute with the remaining rotations on spin~1,
so this ordering is equivalent to that in the figure.
Both fields have magnitude $\Omega$.
The rotation $R_y^{(1)}(\alpha_3)$ takes the first to $-\Omega Z_1$,
then $R_x^{(1)}(\pi)$ reverses its sign;
$R_y^{(2)}(\alpha_4)$ takes the second to $\Omega Z_2$.
The four angles can thus be chosen as
\begin{equation}
\begin{aligned}
\alpha_1&=\atanTwo(b,a),&
\alpha_2&=\pi-\atanTwo(b,c),\\
\alpha_3&=\pi-\atanTwo(r_{bc},a),&
\alpha_4&=\pi-\atanTwo(c,r_{ab}).
\end{aligned}
\label{eq:encoder-angles}
\end{equation}
Here $\atanTwo(y,x)$ is the polar angle of $(x,y)$, with
$\atanTwo(0,0)=0$. These formulas also cover zero couplings;
if all couplings are zero, no gates are needed.
This successive conjugation gives $WH_DW^\dagger=\Omega(Z_1+Z_2)$
on the full Hilbert space, proving Eqs.~\eqref{eq:encoder-diagonal}
and~\eqref{eq:encoder-dm}.

\Needspace{12\baselineskip}
\subsection{Five DM gates}
\label{app:proof}
\label{app:five-angles}
\label{app:sector-proof}

The target evolution and every DM gate conserve the excitation number $N$.
It therefore suffices to establish Eq.~\eqref{eq:identity} separately
in the sectors $N=0,1,2,3$. We first find the angles in the one-excitation
sector. Write $d_{ij}=X_iY_j-Y_iX_j$ and choose the basis
\begin{equation}
|e_1\rangle=|100\rangle,\qquad
|e_2\rangle=|010\rangle,\qquad
|e_3\rangle=|001\rangle.
\label{eq:onebasis}
\end{equation}
Introduce real rotation generators $(L_k)_{mn}=-\epsilon_{kmn}$,
where $x,y,z$ correspond to $1,2,3$ and $\epsilon_{123}=1$.
In this sector,
\begin{equation}
d_{23}=2\ii L_x,\qquad d_{31}=2\ii L_y,\qquad d_{12}=2\ii L_z.
\end{equation}
Thus the matrix of the DM gate $\Ugate{ij}{\theta}$ is a rotation matrix
with angle $2\theta$ about the corresponding axis.
The problem reduces to expressing one three-dimensional rotation
as a sequence of five rotations about fixed axes.

A unit quaternion represents the rotation axis and angle:
a rotation through $2\theta$ about a unit axis $\boldsymbol n$
is represented by $(\cos\theta,\boldsymbol n\sin\theta)$.
For $x_{ij}=tD_{ij}$, with $ij=12,23,31$, define
\begin{equation}
\rho=\sqrt{x_{12}^2+x_{23}^2+x_{31}^2},\qquad
s_\rho=
\begin{cases}
\sin\rho/\rho,&\rho\ne0,\\
1,&\rho=0.
\end{cases}
\label{eq:rho}
\end{equation}
The target exponential has quaternion
\begin{equation}
q_{\rm target}=(q_0,q_x,q_y,q_z)
=\bigl(\cos\rho,\,x_{23}s_\rho,\,x_{31}s_\rho,\,x_{12}s_\rho\bigr).
\label{eq:qtarget}
\end{equation}
For the symmetric sequence in Eq.~\eqref{eq:identity}, multiplying
the quaternions of the three inner gates and then including the two
outer gates gives
\begin{equation}
q_P=\bigl(
\cos C\cos B\cos A,\,
\cos C\sin B,\,
\cos C\cos B\sin A,\,
\sin C
\bigr).
\label{eq:qpal}
\end{equation}
\Needspace{13\baselineskip}
The angles can now be found in order: $C$, then $B$, then $A$.
Matching the last components gives $\sin C=q_z$;
choose $\cos C=\sqrt{q_0^2+q_x^2+q_y^2}\ge0$.
The second component fixes $\cos C\sin B=q_x$.
Choosing also $\cos B\ge0$, the first and third components give
$\cos C\cos B=\sqrt{q_0^2+q_y^2}\equiv w$.
For $w\ne0$, it remains to set $(\cos A,\sin A)=(q_0,q_y)/w$.
Substituting the components of Eq.~\eqref{eq:qtarget} gives the explicit angles
\begin{align}
A&=\atanTwo\!\left(x_{31}s_\rho,\cos\rho\right),\nonumber\\
B&=\atanTwo\!\left(x_{23}s_\rho,
\sqrt{\cos^2\rho+x_{31}^2s_\rho^2}\right),\label{eq:angles-general}\\
C&=\atanTwo\!\left(x_{12}s_\rho,
\sqrt{\cos^2\rho+(x_{23}^2+x_{31}^2)s_\rho^2}\right).\nonumber
\end{align}
This choice ensures $q_P=q_{\rm target}$.
When $w=0$, the terms containing $A$ in Eq.~\eqref{eq:qpal} vanish;
when $\cos C=0$, neither $A$ nor $B$ affects the quaternion.
We may therefore use the convention $\atanTwo(0,0)=0$ in these cases.
The arguments of $C$ cannot both vanish because the target quaternion
has unit norm. The formulas thus apply to all real couplings and times.

It remains to check the other sectors.
The spin flip $F=X_1X_2X_3$ exchanges $N=1$ and $N=2$
and satisfies $Fd_{ij}F=-d_{ij}$. It therefore inverts both the target
evolution and each DM gate. Because the five gates form a symmetric
sequence, replacing every gate by its inverse gives the inverse of
the whole product. Conjugating the established equality by $F$ thus
gives equality of the inverse operators in $N=2$; taking their inverses
establishes the original equality in that sector.
Finally, every $d_{ij}$ annihilates $|000\rangle$ and $|111\rangle$,
so both sides of Eq.~\eqref{eq:identity} are the identity in $N=0,3$.
The result holds on the full Hilbert space.

\Needspace{10\baselineskip}
\subsection{Unequal DM couplings with nonzero exchange}
\label{app:gauge}
\label{app:gauge-local}
\label{app:gauge-longitudinal}

For $J\ne0$, we give explicit local rotations for the reduction
in Sec.~\ref{sec:unequal}:
\begin{equation}
L=\prod_{i=1}^3R_z^{(i)}(\phi_i).
\end{equation}
They leave the $ZZ$ terms unchanged and transform
$z_{ij}=J+\ii D_{ij}$ as
$z'_{ij}=z_{ij}e^{\ii(\phi_j-\phi_i)}$
\cite{Zimboras2013Transport,Essafi2016KagomeMap}.
The phase condition in that section can be written as
\begin{equation}
\operatorname{Re}(z_{12}z_{23}z_{31})
=J\bigl[J^2-D_{12}D_{23}-D_{23}D_{31}-D_{31}D_{12}\bigr]=0.
\label{eq:gaugecriterion}
\end{equation}
For $J\ne0$, this is equivalent to Eq.~\eqref{eq:fluxsurface}.

An explicit choice of angles establishes sufficiency.
Write $r_{ij}=\sqrt{J^2+D_{ij}^2}$ and
$\eta_{ij}=\atanTwo(D_{ij},J)$ for the modulus and argument of $z_{ij}$.
A common rotation of all spins leaves the couplings unchanged,
so we set $\phi_1=0$.
Choose the two remaining angle differences so that
$z'_{12}=\ii r_{12}$ and $z'_{23}=\ii r_{23}$:
\begin{equation}
\phi_1=0,\qquad
\phi_2=\frac\pi2-\eta_{12},\qquad
\phi_3=\pi-\eta_{12}-\eta_{23}.
\label{eq:gaugephases}
\end{equation}
Invariance of the product then gives
\[
z'_{31}=-\frac{z_{12}z_{23}z_{31}}{r_{12}r_{23}}.
\]
This coefficient is purely imaginary by Eq.~\eqref{eq:gaugecriterion},
so all three transverse couplings take the DM form.
Writing $z'_{ij}=\ii\widetilde D_{ij}$, we obtain
$LHL^\dagger=\sum_{(ij)\in\{12,23,31\}}\widetilde D_{ij}d_{ij}+JQ$, where
\begin{equation}
\begin{aligned}
\widetilde D_{12}&=r_{12},\qquad \widetilde D_{23}=r_{23},\\
\widetilde D_{31}
&=\frac{D_{12}D_{23}D_{31}-J^2(D_{12}+D_{23}+D_{31})}{r_{12}r_{23}}.
\end{aligned}
\label{eq:gaugeDM}
\end{equation}
The denominators are nonzero since $J\ne0$.
The formulas use the signed value of $J$ in $\atanTwo(D_{ij},J)$
and apply to either sign of $J$.

The longitudinal operator can be expressed in terms of the excitation
number $N=\sum_i(\Id-Z_i)/2$:
\begin{equation}
Q=Z_1Z_2+Z_2Z_3+Z_3Z_1
=\frac{(3\Id-2N)^2-3\Id}{2}.
\end{equation}
Each $d_{ij}$ transfers an excitation between spins $i$ and $j$
while preserving the total number, so $[\widetilde H_D,N]=0$.
Since $Q$ is a polynomial in $N$, it follows that
$[\widetilde H_D,Q]=0$.
This gives the exact factorization~\eqref{eq:gauge-evolution}
used in Sec.~\ref{sec:unequal}.
Condition~\eqref{eq:gaugecriterion} characterizes the reduction by local
$z$ rotations; it does not exclude other exact circuits outside this family.

\Needspace{12\baselineskip}
\subsection{Derivation of the chain basis change}
\label{app:path}

Left multiplication of $W$ by any unitary commuting with $Z_1+Z_2$
preserves Eq.~\eqref{eq:encoder-diagonal}. Thus the basis change need
not equal the particular circuit in Fig.~\ref{fig:encoder-four}.
The chain circuit in Fig.~\ref{fig:path-circuit}(b) uses this freedom.
Its action follows by conjugating the Pauli terms directly.
Using the notation of Sec.~\ref{sec:path}, immediately after the last
CNOT and before the final local rotations, the Hamiltonian is
\begin{equation}
\begin{aligned}
H_5={}&(a\sin\beta_2+b\cos\beta_2)Z_2Z_3
 -(a\cos\beta_2-b\sin\beta_2)Y_2-cX_2\\
 &+bX_1+(a\sin\beta_1+c\cos\beta_1)Y_1Z_3
 +(a\cos\beta_1-c\sin\beta_1)Z_1\\
={}&(bX_1-r_{ac}Z_1)+(-cX_2-r_{ab}Y_2).
\end{aligned}
\label{eq:path-fields}
\end{equation}
The choices of $\beta_1$ and $\beta_2$ eliminate both nonlocal terms.
The two remaining fields have equal magnitude $\Omega$.
The final rotations $R_y^{(1)}(\beta_3)$ and
$R_y^{(2)}(\beta_4)R_x^{(2)}(\pi/2)$ align both fields along positive $z$.
This proves Eq.~\eqref{eq:path-diagonal} on the full eight-dimensional
space, including cases with zero couplings.

\Needspace{12\baselineskip}
\subsection{Chain circuits with exchange}
\label{app:path-recipes}

\subsubsection{Uniform Heisenberg--DM bonds}
\label{sec:path-uniform}

\begin{figure}[!ht]
\centering
\includegraphics[width=\linewidth]{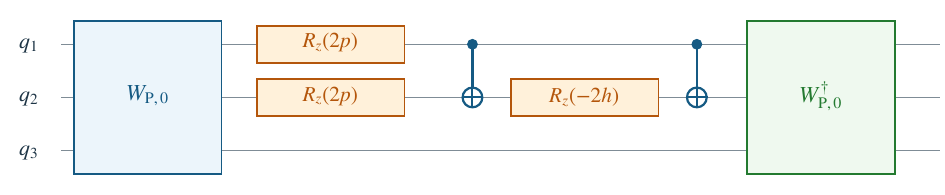}
\caption{Direct chain implementation for uniform $J,D$, with 12 CNOT
 gates. Here $p=\sqrt3Dt$, $h=3Jt$, and $W_{{\rm P},0}$ is the circuit
 of Fig.~\ref{fig:path-circuit} at angles~\eqref{eq:path-equal-angles}.
 Local recompilation reduces the count to 11; the full prescription
 is given in Sec.~S3.2 of the Supplemental Material.}
\label{fig:path-uniform}
\end{figure}

For $D_{12}=D_{23}=D_{31}=D$ and arbitrary uniform exchange $J$, use
$W_{{\rm P},0}=W_{\rm P}(1,1,1)$ from Fig.~\ref{fig:path-circuit}.
Its fixed angles are
\begin{equation}
\begin{gathered}
(\beta_1,\beta_2,\beta_3,\beta_4)=
(3\pi/4,-\pi/4,\pi+\vartheta,-\pi-\vartheta),\\
\vartheta=\atanTwo(1,\sqrt2).
\end{gathered}
\label{eq:path-equal-angles}
\end{equation}
This basis change diagonalizes both $H_0$ and $K$:
$W_{{\rm P},0}KW_{{\rm P},0}^\dagger=-3Z_1Z_2$ by
Eq.~\eqref{eq:joint-diagonal}. Writing $p=\sqrt3Dt$ and $h=3Jt$ gives
\begin{equation}
e^{-\ii t(JK+DH_0)}=W_{{\rm P},0}^\dagger
R_z^{(1)}(2p)R_z^{(2)}(2p)
R_{ZZ}^{(12)}(-2h)W_{{\rm P},0}.
\label{eq:path-uniform}
\end{equation}
Figure~\ref{fig:path-uniform} implements the central $ZZ$ rotation
on the neighboring pair $12$. The formulas apply to either sign of $J$ and $D$;
setting $J=0$ gives the uniform pure-DM circuit.
The local recompilation underlying the bound in
Table~\ref{tab:connectivity-cost} is specified in Sec.~S3.2 of the
Supplemental Material. The resulting 11-CNOT count is a constructive
upper bound; global optimality of the full evolution circuit is not claimed.

\subsubsection{The phase-constrained family}
\label{sec:path-phase}

\begin{figure}[!ht]
\centering
\includegraphics[width=\linewidth]{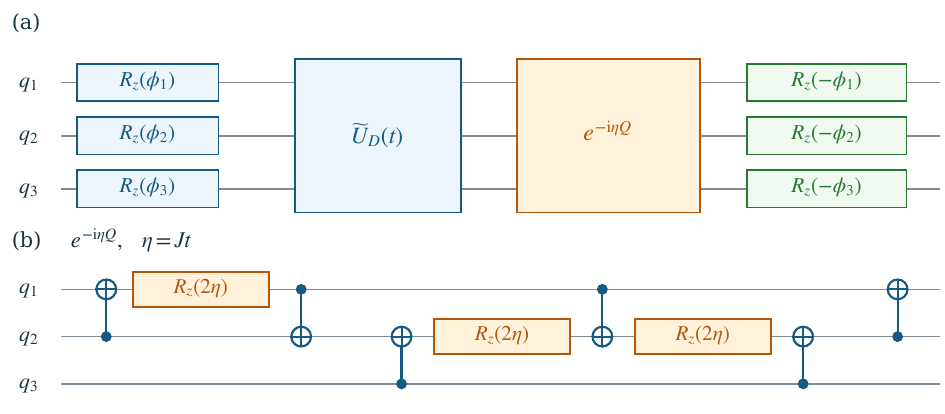}
\caption{Chain circuit for the phase-constrained family.
(a) The full evolution contains at most 16 CNOT gates; the local
rotations use Eq.~\eqref{eq:gaugephases}, and $\widetilde U_D$ is
Eq.~\eqref{eq:path-pure} at couplings~\eqref{eq:gaugeDM}.
(b) The six-CNOT implementation of $e^{-\ii\eta Q}$, where $\eta=Jt$.
All three local rotation angles are $2\eta$.}
\label{fig:path-phase}
\end{figure}

Let the uniform nonzero exchange $J$ and DM couplings satisfy
Eq.~\eqref{eq:fluxsurface}. Set $L=\prod_iR_z^{(i)}(\phi_i)$ using
Eq.~\eqref{eq:gaugephases}, and obtain $\widetilde D_{ij}$ from
Eq.~\eqref{eq:gaugeDM}. The factorization~\eqref{eq:gauge-evolution}
then gives the circuit in Fig.~\ref{fig:path-phase}(a), read from left to right:
apply $L$, evolve under $\widetilde H_D$, apply $e^{-\ii\eta Q}$ with
$\eta=Jt$, and undo $L$. The two central evolutions commute.

Here $\widetilde U_D(t)=e^{-\ii t\widetilde H_D}$ is precisely
Eq.~\eqref{eq:path-pure}, with $W_{\rm P}$ evaluated at the modified
couplings and both central rotation angles equal to
$2t\sqrt{\widetilde D_{12}^2+\widetilde D_{23}^2+\widetilde D_{31}^2}$.
The remaining phase for $Q=Z_1Z_2+Z_2Z_3+Z_3Z_1$ is expanded in
Fig.~\ref{fig:path-phase}(b). The parity circuit implements the phase
associated with the nonadjacent pair $13$ using only gates between
neighboring qubits. The gauge formulas apply to
both signs of $J$; the exchange remains uniform on all three bonds.

\FloatBarrier
\par\smallskip
\Needspace{6\baselineskip}
\begingroup\small
\setlength{\bibsep}{0pt}
\renewcommand{\bibsection}{\par\smallskip\noindent{\normalsize\bfseries\refname}\par\smallskip}

\endgroup

\clearpage
\setcounter{section}{0}
\setcounter{subsection}{0}
\setcounter{equation}{0}
\setcounter{figure}{0}
\setcounter{table}{0}
\renewcommand{\thesection}{S\arabic{section}}
\renewcommand{\thesubsection}{\thesection.\arabic{subsection}}
\renewcommand{\thesubsubsection}{\thesubsection.\arabic{subsubsection}}
\makeatletter\@removefromreset{equation}{section}\makeatother
\renewcommand{\theequation}{S\arabic{equation}}
\renewcommand{\thefigure}{S\arabic{figure}}
\renewcommand{\thetable}{S\arabic{table}}
\renewcommand{\theHsection}{supp.\arabic{section}}
\renewcommand{\theHsubsection}{\theHsection.\arabic{subsection}}
\renewcommand{\theHsubsubsection}{\theHsubsection.\arabic{subsubsection}}
\renewcommand{\theHequation}{supp.\arabic{equation}}
\renewcommand{\theHfigure}{supp.\arabic{figure}}
\renewcommand{\theHtable}{supp.\arabic{table}}
\pdfbookmark[0]{Supplemental Material}{supplement-start}
\begin{center}
{\large\bfseries Supplemental Material for\\[3pt]
Exact quantum circuits for a Heisenberg triangle\\
with Dzyaloshinskii--Moriya interaction\par}
\end{center}
\medskip

Section~S1 derives the analytical transverse-magnetization signals used
in the dynamical example and experiment. Section~S2 specifies the model,
product formula, and error calculation for the kagome comparison.
Section~S3 establishes constructive gate-count bounds on a chain and
gives an explicit local synthesis procedure. Section~S4 provides the
separate lower-bound proof for the basis change with equal DM couplings.
Section~S5 reports the IBM quantum-processor measurement protocol, experimental
error metrics, and results for the individual chains.

Section~S6 describes the coherent-error model and compensation procedure.

\section{Transverse magnetization}

The second dynamical example in the paper uses the triangle Hamiltonian
\begin{equation}
H=J\sum_{(ij)\in\{12,23,31\}}(X_iX_j+Y_iY_j+Z_iZ_j)
 +\sum_{(ij)\in\{12,23,31\}}D_{ij}(X_iY_j-Y_iX_j),
\label{SM-eq:sup-triangle}
\end{equation}
with oriented bonds $12,23,31$, Pauli operators $X_i,Y_i,Z_i$, and $\hbar=1$.
Here $J=1$, $(D_{12},D_{23},D_{31})=(1/2,1,1/3)$, and the initial state is
$|+00\rangle=(|000\rangle+|100\rangle)/\sqrt2$.

Local $z$ rotations transform this Hamiltonian into $\widetilde H_D+Q$,
where $Q=Z_1Z_2+Z_2Z_3+Z_3Z_1$ and $\widetilde H_D$ contains only DM terms
with couplings $(\sqrt5/2,\sqrt2,-\sqrt{10}/3)$. These two operators commute.
The DM evolution has the real return amplitude
\begin{equation}
a_1(t)=\langle100|e^{-\ii t\widetilde H_D}|100\rangle
      =\frac{72+85\cos(\sqrt{157}\,t/3)}{157}.
\label{SM-eq:sup-amplitude}
\end{equation}
The local $z$ rotations do not change this diagonal matrix element.
The eigenvalues of $Q$ are $3$ on $|000\rangle$ and $-1$ throughout the
one-excitation sector. The factor $e^{-\ii tQ}$ therefore gives these
two state components the phases $e^{-3\ii t}$ and $e^{\ii t}$.
Combining them with the return amplitude gives
\begin{equation}
\langle X_1(t)\rangle=a_1(t)\cos4t,\qquad
\langle Y_1(t)\rangle=a_1(t)\sin4t.
\label{SM-eq:sup-magnetization}
\end{equation}
Thus $Q$ contributes $4t$ to the relative phase between excitation
sectors. Populations do not determine this phase, whereas transverse
magnetization is sensitive to it. The argument of the complex signal
$\langle X_1\rangle+\ii\langle Y_1\rangle$ also depends on the sign of
$a_1(t)$: a negative amplitude adds $\pi$, and the phase is undefined
when $a_1(t)=0$.

\section{Kagome calculation}

\subsection{Model and local blocks}

The calculation uses a periodic $2\times2$ kagome cluster of 12 spins.
For an oriented bond, define
$h_{ij}=J(X_iX_j+Y_iY_j+Z_iZ_j)+D(X_iY_j-Y_iX_j)$.
The upward and downward triangles are the directed cycles
\begin{equation}
\begin{aligned}
\mathcal T_{\triup}&=\{(1,2,3),(4,5,6),(7,8,9),(10,11,12)\},\\
\mathcal T_{\tridown}&=\{(1,5,9),(4,2,12),(7,11,3),(10,8,6)\}.
\end{aligned}
\label{SM-eq:sup-triangles}
\end{equation}
Each cycle $(i,j,k)$ contributes $h_{ij}+h_{jk}+h_{ki}$.
Let $A$ and $B$ be the sums over $\mathcal T_{\triup}$ and
$\mathcal T_{\tridown}$, respectively. We study evolution under $H=A+\varepsilon B$ with
$J=1$, $D=0.2$, for $(\varepsilon,T)=(0.2,5)$ and $(1,1)$.
Thus the weaker downward coupling is accompanied by a longer time;
both cases have $\varepsilon JT=1$.

The triangle decomposition has two layers, $H_1=A$ and
$H_2=\varepsilon B$, each containing four disjoint triangles.
The bond decomposition uses four layers of six disjoint bonds:
\begin{equation}
\begin{aligned}
\mathcal E_1&=\{(1,2),(6,4),(7,8),(12,10),(5,9),(11,3)\},\\
\mathcal E_2&=\{(2,3),(8,9),(1,5),(12,4),(7,11),(6,10)\},\\
\mathcal E_3&=\{(3,1),(4,5),(9,7),(10,11),(2,12),(8,6)\},\\
\mathcal E_4&=\{(5,6),(11,12),(9,1),(4,2),(3,7),(10,8)\}.
\end{aligned}
\label{SM-eq:sup-bonds}
\end{equation}
Each bond-layer Hamiltonian is the sum of the corresponding bond terms, weighted by
$1$ or $\varepsilon$ according to the triangle family in
Eq.~\eqref{SM-eq:sup-triangles}. These definitions fix all periodic bonds
and DM orientations without reference to a drawing.

\subsection{Product formula and resource count}

Both decompositions use the same fourth-order Omelyan formula.
Let $P$ and $Q$ denote the two layer Hamiltonians, with
$E_P(s)=e^{-\ii sP}$ and similarly for $Q$. Then
\begin{equation}
\begin{aligned}
S_4(\delta)={}&E_P(a_1\delta)E_Q(b_1\delta)
 E_P(a_2\delta)E_Q(b_2\delta)E_P(a_3\delta)\\
 &\times E_Q(b_2\delta)E_P(a_2\delta)
 E_Q(b_1\delta)E_P(a_1\delta),
\end{aligned}
\label{SM-eq:sup-omelyan}
\end{equation}
where $\delta=T/r$. The coefficients used are those of Appendix C.1.2
in the preprint of Bosse et al.~\cite{SM-Bosse2025}:
\begin{equation}
\begin{aligned}
a_1&=0.5316386245813512,& a_2&=-0.3086019704406066,\\
b_1&=-0.04375142191737413,& b_2&=\tfrac12-b_1,\\
a_3&=1-2(a_1+a_2).
\end{aligned}
\label{SM-eq:sup-coefficients}
\end{equation}
For $m$ layers, set
\begin{equation}
\begin{gathered}
(\alpha_1,\ldots,\alpha_4)=(a_1,a_2,a_3,a_2),\qquad
(\beta_1,\ldots,\beta_4)=(b_1,b_2,b_2,b_1),\\
d_0=0,\qquad c_k=\alpha_k-d_{k-1},\qquad d_k=\beta_k-c_k.
\end{gathered}
\label{SM-eq:sup-sweeps}
\end{equation}
For each $k=1,\ldots,4$, apply the following forward and reverse sweeps
in the listed chronological order, with $E_j(s)=e^{-\ii sH_j}$:
\begin{equation}
E_{\sigma(1)}(c_k\delta),\ldots,E_{\sigma(m)}(c_k\delta),\quad
E_{\sigma(m)}(d_k\delta),\ldots,E_{\sigma(1)}(d_k\delta).
\label{SM-eq:sup-sweep-order}
\end{equation}
Here $\sigma$ specifies the layer order. At $m=2$, combining adjacent
identical layers gives Eq.~\eqref{SM-eq:sup-omelyan}. After this combination
within and between $r$ steps, there are $8r(m-1)+1$ layers.
A triangle layer costs $4\times10$ CNOT gates, and a bond layer
$6\times3$~\cite{SM-VidalDawson2004}, so
\begin{equation}
N_{\rm CNOT}^{\rm tri}=40(8r+1),\qquad
N_{\rm CNOT}^{\rm bond}=18(24r+1).
\label{SM-eq:sup-costs}
\end{equation}
Negative substep durations are implemented by changing rotation angles
in either decomposition. The counts describe these logical circuits;
no routing or hardware noise is included.

\subsection{Error and numerical examples}

For the circuit $C$ and the exact propagator $U=e^{-\ii TH}$, the error is
\begin{equation}
E_{\max}=\max_{\|\psi\|=1}\left[1-
 |\langle\psi|U^\dagger C|\psi\rangle|^2\right].
\label{SM-eq:sup-error}
\end{equation}
Let $e^{\ii\theta_j}$ be the eigenvalues of $V=U^\dagger C$.
The possible overlaps $\langle\psi|V|\psi\rangle$ form their convex hull.
Its distance from the origin gives the minimum overlap modulus.
If $w$ is the width of the shortest circular arc containing all eigenvalues,
\begin{equation}
E_{\max}=\begin{cases}\sin^2(w/2),&w<\pi,\\1,&w\ge\pi.\end{cases}
\label{SM-eq:sup-arc}
\end{equation}
We obtain $w$ as $2\pi$ minus the largest gap between sorted eigenphases,
including the wraparound gap. The calculation uses the 13 excitation
sectors: exact sector propagators follow from diagonalization, while the
circuit is formed from its one-step matrix raised to power $r$.
All 4096 eigenvalues of $V$ are pooled before evaluating
Eq.~\eqref{SM-eq:sup-arc}. This retains relative phases between sectors.

Table~\ref{SM-tab:sup-cases} gives concrete circuits reaching three target
errors. The orders were selected among the two triangle orders and the
24 permutations of the fixed bond layers. The table includes the achieved
errors, and the listed parameters fully specify the circuits. These are
examples meeting the target, not certificates of a globally minimal
step or gate count.

\begin{table}[H]
\centering\small
\setlength{\tabcolsep}{4pt}
\caption{Circuits with $E_{\max}\le\eta$. The layer order $\sigma$ is
written with the labels defined above; for example, $4312$ means
$(\mathcal E_4,\mathcal E_3,\mathcal E_1,\mathcal E_2)$.}
\label{SM-tab:sup-cases}
\begin{tabular}{@{}cccrrrcrrr@{}}
\hline
$(\varepsilon,T)$&$\eta$&\multicolumn{4}{c}{Triangles}&\multicolumn{4}{c}{Bonds}\\
&&$\sigma$&$r$&CNOT&$E_{\max}$&$\sigma$&$r$&CNOT&$E_{\max}$\\\hline
$(0.2,5)$&$10^{-2}$&21&13&4200&$3.68\times10^{-3}$&4312&17&7362&$7.26\times10^{-3}$\\
&$10^{-3}$&21&16&5160&$9.35\times10^{-4}$&4312&22&9522&$9.96\times10^{-4}$\\
&$10^{-4}$&21&18&5800&$8.48\times10^{-5}$&4312&30&12978&$8.83\times10^{-5}$\\
$(1,1)$&$10^{-2}$&21&5&1640&$4.00\times10^{-3}$&1243&5&2178&$5.69\times10^{-3}$\\
&$10^{-3}$&21&6&1960&$6.49\times10^{-4}$&3421&7&3042&$3.88\times10^{-4}$\\
&$10^{-4}$&12&8&2600&$4.86\times10^{-5}$&4312&9&3906&$5.46\times10^{-5}$\\\hline
\end{tabular}
\end{table}

The comparison concerns the choice of local blocks within the same
fourth-order formula. The lower CNOT cost in these examples demonstrates the benefit of
exact triangle blocks; the reduction depends on the regime and target error.

\subsection{Asymptotic cost balance}

The benefit of an exact local block is determined jointly by its synthesis
cost and the error of the resulting product formula. To make this balance
explicit, consider two decompositions of the same order $p$ and suppose
their operator-norm errors satisfy
$\delta_r=\|U_r(T)-U(T)\|\simeq\kappa_p(T)r^{-p}$, with nonzero leading
coefficients. If their two-qubit gate counts are $G(r)=cr+O(1)$, the
leading cost at a common small tolerance is proportional to
$c\kappa_p(T)^{1/p}$. Exact blocks are therefore favored in this regime when
\begin{equation}
\frac{c_{\rm block}}{c_{\rm bond}}
 < \left(\frac{\kappa_{p,\rm bond}(T)}
 {\kappa_{p,\rm block}(T)}\right)^{1/p}.
\label{SM-eq:block-cost-balance}
\end{equation}
Here $c$ counts an entire repeated product-formula step after gate merging,
including routing if required, rather than a single local exponential.
Nested commutators control the error coefficients
\cite{SM-Childs2021Trotter}; weak couplings between exact blocks can reduce
them, but are not required for a gain. This asymptotic estimate does not
replace comparison at integer step counts. The error $\delta_r$ is also
distinct from the worst-case infidelity $E_{\max}$ used in
Table~\ref{SM-tab:sup-cases}.

\section{Gate counts for circuits on a nearest-neighbor chain}
\label{SM-sec:sup-resources}

This section establishes the constructive bounds in Table~1 of the
main text. For uniform couplings, local synthesis of the central
two-qubit fragment reduces the complete circuit from 12 to 11 CNOT
gates; Section~\ref{SM-sec:sup-middle-synthesis} gives the explicit recipe.
The outer circuit fragments remain fixed, so this improvement does
not establish a global minimum for the three-qubit evolution.

All two-qubit gates in this section act between neighbors in the chain
$1$--$2$--$3$.
We use ordinary Pauli matrices and
$R_\alpha(\varphi)=\exp(-\ii\varphi\sigma_\alpha/2)$.
We write $\mathrm{CX}_{i\to j}$ for CNOT with control $i$ and target $j$.
In matrix products, the rightmost factor acts first; explicitly listed
gate sequences are given in chronological order. Let
\begin{equation}
\begin{aligned}
H_D&=\sum_{(ij)\in\{12,23,31\}}D_{ij}(X_iY_j-Y_iX_j),\\
H_0&=H_D\big|_{D_{12}=D_{23}=D_{31}=1},\qquad
K=\sum_{(ij)\in\{12,23,31\}}(X_iX_j+Y_iY_j+Z_iZ_j),\\
Q&=Z_1Z_2+Z_2Z_3+Z_3Z_1,\qquad
\Omega=\sqrt{D_{12}^2+D_{23}^2+D_{31}^2}.
\end{aligned}
\label{SM-eq:sup-operators}
\end{equation}
The five-CNOT circuit $W_{\rm P}$ and its four angles are specified in
Sec. 5 of the main text; Appendix A.4 gives a short proof. It obeys
$W_{\rm P}H_DW_{\rm P}^\dagger=\Omega(Z_1+Z_2)$ for arbitrary real
couplings. We use its fixed specialization
$W_{{\rm P},0}=W_{\rm P}(1,1,1)$ for uniform Heisenberg--DM couplings.
Since $K=3I-H_0^2/2$, the same circuit gives
$W_{{\rm P},0}KW_{{\rm P},0}^\dagger=-3Z_1Z_2$.

For arbitrary pure DM couplings,
\begin{equation}
e^{-\ii tH_D}=W_{\rm P}^\dagger
\bigl[R_z^{(1)}(2\Omega t)R_z^{(2)}(2\Omega t)\bigr]W_{\rm P}
\label{SM-eq:sup-pure-DM}
\end{equation}
uses at most 10 CNOT gates.

\subsection{Local compression of the uniform evolution}
For the uniform Heisenberg--DM family, $D_{12}=D_{23}=D_{31}=D$,
write $W_{{\rm P},0}=\mathcal L_5\mathrm{CX}_{1\to2}F$.
The prefix $F$ contains the first four CNOT gates and all local
operations preceding the last CNOT, including the
$R_z^{(2)}(\pi/2)$ immediately before it.
The final local layer is explicitly
\begin{equation}
\mathcal L_5=R_y^{(1)}(\vartheta+\pi)
 R_y^{(2)}(-\pi-\vartheta)R_x^{(2)}(\pi/2),\qquad
\vartheta=\atanTwo(1,\sqrt2).
\label{SM-eq:sup-last-layer}
\end{equation}
Since $\mathcal L_5$ acts only on qubits 1 and 2,
the full evolution has the form
\begin{equation}
\begin{aligned}
e^{-\ii t(JK+DH_0)}&=F^\dagger M_{12}(t)F,\\
M_{12}(t)&=\mathrm{CX}_{1\to2}\mathcal L_5^\dagger
 e^{-\ii\sqrt3Dt(Z_1+Z_2)+\ii3JtZ_1Z_2}
 \mathcal L_5\mathrm{CX}_{1\to2}.
\end{aligned}
\label{SM-eq:path-middle}
\end{equation}
The operator $M_{12}(t)$ acts on the adjacent pair $12$. Its direct
implementation contains four CNOT gates, including two for the central
$ZZ$ rotation. Exact two-qubit synthesis needs at most three
\cite{SM-VatanWilliams2004TwoQubit}, giving at most $4+3+4=11$ CNOT gates
for the complete circuit on the chain, for any real $J,D,t$.
Section~\ref{SM-sec:sup-middle-synthesis} gives the explicit three-CNOT
sequence and all its local parameters.

For this fixed basis change, three CNOT gates are generically necessary for
the middle fragment itself. Regard $M_{12}$ as a matrix in $SU(4)$ and define
$\gamma(M_{12})=M_{12}(Y\otimes Y)M_{12}^{\mathsf T}(Y\otimes Y)$.
Here $\mathsf T$ is the transpose, not the adjoint.
Using the local layer in Eq.~\eqref{SM-eq:sup-last-layer} gives
\begin{equation}
\operatorname{Im}\Tr\gamma(M_{12})
 =-\frac89\sin^2(2\sqrt3Dt)\sin(6Jt).
\label{SM-eq:path-middle-invariant}
\end{equation}
For example, this follows by writing $p=\sqrt3Dt$, $h=3Jt$,
$u=\cos^2p\cos h-\ii\sin^2p\sin h$ and
$v=-\sin^2p\cos h+\ii\cos^2p\sin h$; Pauli orthogonality gives
$\Tr\gamma(M_{12})=4(u^2-v^2/9)$.
A two-qubit unitary can be implemented with at most two CNOT gates
if and only if this trace is real \cite[Proposition III.3]{SM-Shende2004Recognizing}.
A nonzero right-hand side of Eq.~\eqref{SM-eq:path-middle-invariant}
therefore proves the three-CNOT requirement.
Thus 11 is tight when the eight outer CNOT gates are fixed and only
$M_{12}(t)$ is resynthesized. It is not a lower bound for the full
three-qubit evolution; changing a larger circuit fragment or the
basis change is not excluded. At exceptional parameters the middle
fragment can require fewer gates.

\subsection{Explicit synthesis of the middle fragment}
\label{SM-sec:sup-middle-synthesis}

We give a finite procedure for computing the parameters of the three-CNOT replacement,
so that Eq.~\eqref{SM-eq:path-middle} can be implemented without a circuit
optimizer. All matrices below use the basis $|q_1q_2\rangle$, with qubit 1
as the most significant bit. Put $p=\sqrt3Dt$, $h=3Jt$ and
$C=\mathrm{CX}_{1\to2}$. The required two-qubit matrix is
\begin{equation}
\begin{gathered}
M=C\mathcal L_5^\dagger E(p,h)\mathcal L_5 C,\\
E(p,h)=\operatorname{diag}
\bigl(e^{\ii(h-2p)},e^{-\ii h},e^{-\ii h},e^{\ii(h+2p)}\bigr),
\end{gathered}
\label{SM-eq:sup-middle-matrix}
\end{equation}
where $\mathcal L_5$ is given in Eq.~\eqref{SM-eq:sup-last-layer}.
In particular, $M\in SU(4)$.

We first obtain its exact Cartan decomposition with local factors
\begin{equation}
M=(K_{L1}\otimes K_{L2})
 e^{\ii(aX\otimes X+bY\otimes Y+cZ\otimes Z)}
 (K_{R1}\otimes K_{R2}),
\label{SM-eq:sup-middle-cartan}
\end{equation}
with four $SU(2)$ matrices and real parameters $a,b,c$.
These parameters need not lie in a Weyl chamber. The following procedure
fixes all phases needed for the matrix equality.

\begin{enumerate}
\item Introduce the magic-basis matrix
\begin{equation}
B=\frac1{\sqrt2}
\begin{pmatrix}
1&\ii&0&0\\
0&0&\ii&1\\
0&0&\ii&-1\\
1&-\ii&0&0
\end{pmatrix},\qquad
U_B=B^\dagger MB,\qquad S=U_B^{\mathsf T}U_B.
\label{SM-eq:sup-magic}
\end{equation}
The real and imaginary parts of $S$ are real symmetric matrices.
They commute because $S$ is unitary.

\item Choose a simultaneous real orthonormal eigenbasis $V$ of those
two matrices. Change the sign of one column if necessary so that
$\det V=1$. Then set
\begin{equation}
\delta_j=\tfrac12\arg\bigl[(V^{\mathsf T}SV)_{jj}\bigr],\qquad
O_L=U_BV\operatorname{diag}(e^{-\ii\delta_j}).
\label{SM-eq:sup-magic-phases}
\end{equation}
Here $\arg$ is the principal argument. The matrix $O_L$ is real orthogonal. If $\det O_L=-1$, add $\pi$ to
$\delta_4$ and recompute $O_L$. Both $O_L$ and $V$ then belong to
$SO(4)$. Their determinants and $\det U_B=1$ imply
$\sum_j\delta_j=2\pi n$ for an integer $n$. Replace
$\delta_4$ by $\delta_4-2\pi n$, leaving $O_L$ unchanged, to obtain
$\sum_j\delta_j=0$.

\item Transform the orthogonal factors back:
\begin{equation}
BO_LB^\dagger=K_{L1}\otimes K_{L2},\qquad
BV^{\mathsf T}B^\dagger=K_{R1}\otimes K_{R2}.
\label{SM-eq:sup-local-factors}
\end{equation}
The $SO(4)$--$SU(2)\otimes SU(2)$ correspondence ensures the local
factorization; an explicit extraction is given below. The diagonal
factor in the magic basis gives
\begin{equation}
\begin{aligned}
a&=(\delta_1-\delta_2+\delta_3-\delta_4)/4,\\
b&=(-\delta_1+\delta_2+\delta_3-\delta_4)/4,\\
c&=(\delta_1+\delta_2-\delta_3-\delta_4)/4.
\end{aligned}
\label{SM-eq:sup-cartan-angles}
\end{equation}
\end{enumerate}

Let $T$ denote either matrix on the left-hand side of
Eq.~\eqref{SM-eq:sup-local-factors}. Its $SU(2)$ factors can be extracted
by rank-one realignment. Form
$\mathcal R_{(i,k),(j,l)}=T_{(i,j),(k,l)}$, with all four binary indices
in $\{0,1\}$. Choose an entry $\mathcal R_{rs}$ of maximum absolute value.
Reshape column $s$ into a $2\times2$ matrix $A_0$, and reshape row $r$
divided by $\mathcal R_{rs}$ into $B_0$; then $T=A_0\otimes B_0$.
Normalize by
\begin{equation}
q=\sqrt{\tfrac12\Tr(A_0^\dagger A_0)},\quad
A=A_0/q,\quad B_1=qB_0,\quad z=\sqrt{\det A}.
\label{SM-eq:sup-local-extraction}
\end{equation}
The desired $SU(2)$ factors are $A/z$ and $zB_1$.
Either square-root choice gives the same tensor product.

\begin{figure}[H]
\centering
\includegraphics[width=\linewidth]{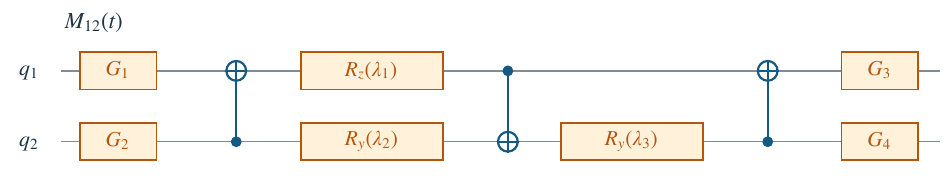}
\caption{Local replacement of the central fragment.
The gates $G_k$ and angles $\lambda_j$ are specified in
Eq.~\eqref{SM-eq:sup-middle-gates}. The circuit uses three CNOT gates;
its matrix equals $e^{-\ii\pi/4}M$.}
\label{SM-fig:sup-middle}
\end{figure}

The standard three-CNOT realization of the Cartan factor
\cite{SM-VatanWilliams2004TwoQubit} now becomes fully explicit. Define
\begin{equation}
\begin{aligned}
\lambda_1&=\pi/2-2c,&
\lambda_2&=2a-\pi/2,&
\lambda_3&=\pi/2-2b,\\
G_1&=K_{R1},&G_2&=R_z(-\pi/2)K_{R2},\\
G_3&=K_{L1}R_z(\pi/2),&G_4&=K_{L2}.
\end{aligned}
\label{SM-eq:sup-middle-gates}
\end{equation}
Figure~\ref{SM-fig:sup-middle} shows the resulting two-qubit circuit.

Read the following sequence chronologically, from left to right and then
from top to bottom:
\begin{equation}
\begin{array}{lll}
G_1\otimes G_2,&\mathrm{CX}_{2\to1},&
R_z^{(1)}(\lambda_1)R_y^{(2)}(\lambda_2),\\
\mathrm{CX}_{1\to2},&R_y^{(2)}(\lambda_3),&\mathrm{CX}_{2\to1},\\
G_3\otimes G_4.&&
\end{array}
\label{SM-eq:sup-middle-sequence}
\end{equation}
If $V_3$ is its matrix, then $M=e^{\ii\pi/4}V_3$.
The parameter-independent global phase has no effect on state dynamics;
it is retained in the verification code.

Each local gate can be expanded into elementary rotations as well.
For $G\in SU(2)$, put $u=G_{00}$, $v=-G_{01}$ and set
\begin{equation}
\alpha=-\arg u-\arg v,\qquad
\beta=2\atanTwo(|v|,|u|),\qquad
\gamma=-\arg u+\arg v.
\label{SM-eq:sup-local-zyz}
\end{equation}
Then $G=R_z(\alpha)R_y(\beta)R_z(\gamma)$, so the chronological
angles are $\gamma,\beta,\alpha$. The convention $\arg0=0$ also covers zero matrix entries.

Repeated eigenvalues introduce no exceptional circuit branch: any real
simultaneous eigenbasis gives the same $M$ after reconstruction.
For numerical stability, the NumPy implementation
\texttt{middle3\_analytic.py} tests the seven fixed combinations
$\operatorname{Re}S+\eta\operatorname{Im}S$, with
$\eta\in\{0,1,-1,2,-2,\sqrt2,-\sqrt2\}$, resolves degenerate subspaces
using $\operatorname{Im}S$, and selects the basis with the smallest
simultaneous-diagonalization residual. This is a finite sequence of
$4\times4$ real symmetric eigenproblems, not an optimization of gate
parameters. The function \texttt{three\_cnot\_gates(p,h)} returns the
chronological elementary gates and the global phase.
The function \texttt{uniform\_path\_gates(J,D,t)} adds the four-CNOT
prefix and its inverse, returning the complete 11-CNOT circuit.
\FloatBarrier

\subsection{The phase-constrained family}
Consider the Hamiltonian in Eq.~\eqref{SM-eq:sup-triangle}, with the same
nonzero exchange $J$ on every bond, and impose
\begin{equation}
J^2=D_{12}D_{23}+D_{23}D_{31}+D_{31}D_{12}.
\label{SM-eq:sup-phase-condition}
\end{equation}
Local $z$ rotations $L$ give $LHL^\dagger=\widetilde H_D+JQ$,
where $\widetilde H_D$ is a pure DM Hamiltonian and
$[\widetilde H_D,Q]=0$ (main text, Sec. 2.3 and Appendix A.3).
Consequently,
\begin{equation}
e^{-\ii tH}=L^\dagger e^{-\ii t\widetilde H_D}e^{-\ii JtQ}L.
\label{SM-eq:sup-phase-evolution}
\end{equation}
The factor $e^{-\ii t\widetilde H_D}$ uses the arbitrary-coupling
basis change on the chain and requires at most 10~CNOT gates.
The remaining factor
$e^{-\ii\theta Q}$, with $\theta=Jt$, has the following chronological
sequence, read from left to right and then from top to bottom:
\begin{equation}
\begin{array}{llll}
\mathrm{CX}_{2\to1},&R_z^{(1)}(2\theta),&
\mathrm{CX}_{1\to2},&\mathrm{CX}_{3\to2},\\
R_z^{(2)}(2\theta),&\mathrm{CX}_{1\to2},&
R_z^{(2)}(2\theta),&\mathrm{CX}_{3\to2},\\
\mathrm{CX}_{2\to1}.&&&
\end{array}
\label{SM-eq:path-Q-parity}
\end{equation}
For an input computational basis state $|x_1x_2x_3\rangle$, the three
rotations act on the parities $x_1\oplus x_2$, $x_1\oplus x_3$ and
$x_2\oplus x_3$, respectively. The final gates restore all three bits.
The acquired phase is therefore exactly
$\exp[-\ii\theta\sum_{(ij)\in\{12,23,31\}}(-1)^{x_i+x_j}]$.
All six CNOT gates act on adjacent pairs, so the full circuit
costs at most 16 CNOT gates. The construction includes both
signs of the nonzero $J$ satisfying Eq.~\eqref{SM-eq:sup-phase-condition}.
The outer gauge rotations $L,L^\dagger$ add no two-qubit gates.

\Needspace{12\baselineskip}
\section{Minimality of the basis change on a chain}
\label{SM-app:minimality}
\label{SM-sec:sup-minimality}
\providecommand{\minE}{\mathcal E}
\providecommand{\minsig}{\boldsymbol\sigma}
\providecommand{\minspan}{\operatorname{span}}
\providecommand{\mincof}{\operatorname{cof}}
\providecommand{\minCZ}{\mathrm{CZ}}
\newtheorem{minpaththm}{Theorem}
\renewcommand{\theminpaththm}{S\arabic{minpaththm}}

We prove the lower bound for equal unit DM couplings. Unlike the
explicit construction in Sec. 5 of the main text, this bound applies to every
admissible diagonalizer, without fixing its local angles or its action
within degenerate eigenspaces.

For the proof we use the locally equivalent controlled-$Z$ gate,
\[
\mathrm{CZ}_{ij}=H_j\mathrm{CX}_{i\to j}H_j,
\]
where $H_j$ is the Hadamard gate on the target qubit and
$\mathrm{CX}_{i\to j}$ denotes CNOT. Each substitution preserves the
edge and the number of two-qubit gates. The Hadamard gates can be
absorbed into the arbitrary local layers; their phases are irrelevant
to the conjugation condition. Thus the minimum CNOT and CZ counts coincide.

\subsection{Statement and notation}
All identities below act on the full eight-dimensional space.
We consider unitary circuits on these three qubits, without ancillary
qubits or intermediate measurements. Local gates are arbitrary
single-qubit unitaries. With ordinary Pauli matrices, let
\begin{align}
H_0&=(X_1Y_2-Y_1X_2)+(X_2Y_3-Y_2X_3)+(X_3Y_1-Y_3X_1),\\
T&=\sqrt3(Z_1+Z_2).
\end{align}
We call qubits 1 and 2 \emph{active} and qubit 3 \emph{inactive}.
These terms refer only to the fields in the target operator $T$.
Let $\mathcal C_m$ denote circuits of single-qubit unitaries and CNOT
gates along the chain whose middle logical qubit is $m$.

\begin{minpaththm}\label{SM-thm:path-minimality}
Any circuit of arbitrary local gates and CNOT gates on a
three-qubit nearest-neighbor chain that satisfies $WH_0W^\dagger=T$ contains at least
five CNOT gates. The same statement holds with CZ as the two-qubit gate.
The lower bound holds for each choice $m=1,2,3$ of
the middle logical qubit. A five-CNOT circuit exists on the chain
$1$--$2$--$3$. Consequently,
\[
\min_{m\in\{1,2,3\}}\ \min_{\substack{W\in\mathcal C_m\\WH_0W^\dagger=T}}
N_{\mathrm{CNOT}}(W)=5.
\]
\end{minpaththm}
The construction establishes the upper bound for $m=2$. Without
an additional construction, the theorem does not assert attainability
with five gates for each fixed placement.

We work with the inverse circuit $U=W^\dagger$, so that
$UTU^\dagger=H_0$. Inversion preserves the CZ count and leaves every
local layer arbitrary. Gate words below are written chronologically,
in their order of action on $T$ or $Z_1Z_2$.

For a qubit $a$, write an operator as
\[
F=I_a\otimes F_0+\sum_{i=x,y,z}\sigma_a^i\otimes F_i,
\qquad \minE_a(F)=\tfrac12\Tr_aF=F_0,
\]
and set $F_v=\sum_i v_iF_i$ for $v\in\mathbb R^3$.
A local rotation of $a$ mixes the three coefficients $F_i$ by a real
orthogonal matrix. A unitary on the other two qubits conjugates them
simultaneously. These operations preserve the dimension of their
span, the dimension of their common kernel, and the possible spectra
of their real linear combinations.

The pair-interaction matrix for qubits $a,b$ is the real matrix $J$
in the term $\sum_{ij}J_{ij}\sigma_a^i\sigma_b^j$, with identity on
the third qubit. Local rotations act as $J\mapsto R_aJR_b^T$ and
therefore preserve its rank. This matrix notation is used only within
the proof; its entries are not the uniform exchange parameter.

\paragraph{Proof outline.}
We first use the two cuts of the chain to require at least two CZ
gates on each edge, leaving six words within a four-gate budget.
The spectra of operator coefficients exclude the grouped words.
For alternating words we use a necessary condition from the squared
Hamiltonian; for nested words we use invariants of the centered
coefficients of the Hamiltonian itself. Once all six words are
excluded, the explicit circuit in the main text establishes
attainability with five CNOT gates.

\subsection{Each edge requires at least two CZ gates}
Consider a cut separating a leaf $a$ from the remaining qubits $b,c$.
If at most one CZ crosses the cut, we may combine all other operations
into arbitrary unitaries on the two sides before and after that CZ.
Enlarging the allowed class this way can only make realization easier.

If $a$ is active, immediately before the CZ the source has the form
\[
A_a\otimes I_{bc}+I_a\otimes B_{bc},\qquad
\|B_{bc}\|_{\mathrm{op}}=\sqrt3.
\]
Here $A_a$ is a rotated Pauli matrix times $\sqrt3$, and $B_{bc}$
is unitarily equivalent to the remaining active field. Conjugation by
CZ creates no identity component on $a$ from the first term. For the
second term,
\[
\minE_a\!\left(\minCZ_{ab}(I_a\otimes B)\minCZ_{ab}\right)
=\frac{B+Z_bBZ_b}{2},\qquad
\left\|\frac{B+Z_bBZ_b}{2}\right\|_{\mathrm{op}}\le\sqrt3.
\]
Subsequent unitaries on $bc$ preserve this norm. However,
\[
\minE_a(H_0)=\pm(X_bY_c-Y_bX_c),\qquad
\|\minE_a(H_0)\|_{\mathrm{op}}=2,
\]
since a two-qubit DM term has spectrum $(-2,0,0,2)$. This is a
contradiction. The same bound applies when no CZ crosses the cut.

If $a$ is inactive, the source is $I_a\otimes B$. One CZ creates
at most the two directions $I_a,Z_a$, giving operator Schmidt rank
at most two across the cut. In contrast, $H_0$ has rank three:
with cyclic labels,
\[
H_0=I_a\otimes(X_bY_c-Y_bX_c)
+X_a\otimes(Y_b-Y_c)+Y_a\otimes(X_c-X_b),
\]
whose three coefficients are linearly independent.

Thus a circuit with at most four CZ gates must contain exactly two
on each edge. Calling the two edges $A,B$, the only remaining words are
\[
AABB,\quad BBAA,\quad ABAB,\quad BABA,\quad ABBA,\quad BAAB.
\]
Repeated letters have not been cancelled: every intervening local
layer remains arbitrary.

\subsection{Grouped words: AABB and BBAA}
Let $A=ab$ and $B=bc$, and replace each pair of CZ gates and its
local layers by an arbitrary two-qubit unitary. After the first block
on $ab$, every nonidentity coefficient on $a$ has the form
$M_b\otimes I_c$. The second block conjugates all these coefficients
by the same unitary on $bc$. Hence every real linear combination of
them has eigenvalues of even multiplicity.

For $H_0$, every nonzero real linear combination of the nonidentity
coefficients on a leaf has the form
\[
u(Y_b-Y_c)+v(X_c-X_b),
\]
with spectrum $(-2r,0,0,2r)$, where $r=\sqrt{u^2+v^2}>0$.
Its two nonzero eigenvalues are simple, which is impossible for the
grouped circuit. Interchanging the edge names excludes the other
grouped word.

\subsection{A necessary condition from the squared Hamiltonian}
Pauli multiplication gives
\begin{align}
H_0^2&=6I-2\sum_{i<j}(X_iX_j+Y_iY_j+Z_iZ_j),\\
T^2&=6I+6Z_1Z_2.
\end{align}
Every candidate $U$ must therefore satisfy
\begin{equation}\label{SM-eq:min-square}
U(Z_1Z_2)U^\dagger=\mathcal R,\qquad
\mathcal R:=\frac{H_0^2}{6}-I
=-\frac13\sum_{i<j}\minsig_i\cdot\minsig_j.
\end{equation}
All three pair-interaction matrices of $\mathcal R$ have rank three.
Its nonidentity coefficients on any leaf $a$ are
\[
\mathcal R_i=-\frac13(\sigma_b^i+\sigma_c^i),\qquad i=x,y,z.
\]
They are linearly independent and have a nonzero common kernel:
the singlet $|s\rangle_{bc}=(|01\rangle-|10\rangle)/\sqrt2$
obeys $\mathcal R_i|s\rangle=0$ for all three directions.
Together with $\mathcal R_0=-\minsig_b\cdot\minsig_c/3$, the
coefficients are independent, so the operator Schmidt rank is four.

\subsection{Alternating words with an inactive middle qubit}
Write the path as $a$--$m$--$b$, with $m$ inactive, and let
$A=am$, $B=mb$. The source in Eq.~\eqref{SM-eq:min-square} is a
product of Pauli matrices on the leaves. It suffices to examine
$ABAB$: exchanging the leaves covers the other word.

After the first $A$, $\minE_a(F)=0$. Indeed, a Pauli matrix on $a$
remains nonidentity on $a$ under this CZ, while the middle qubit
initially carries $I_m$. The second gate $B$ does not act on $a$.
Consequently $\minE_a(F)=0$ just before the third CZ as well.

The third gate $A$ can create an identity coefficient on $a$ only
from the components $Z_aX_m$ and $Z_aY_m$. In its CZ axes this gives
\[
K:=\minE_a(F)=X_m\otimes C_b+Y_m\otimes D_b.
\]
Thus $\minE_m(K)=0$, and the span of the three nonidentity
coefficients on $m$ has dimension at most two. Local layers preserve
both properties.

The last gate $B$ must map $K$ to $\minE_a(\mathcal R)$, a pure
pair interaction with an invertible interaction matrix $J$. Absorb
the final local rotations into $J$. Before the last CZ,
\[
K=\minCZ_{mb}\Bigl(\sum_{u,v}J_{uv}\sigma_m^u\sigma_b^v\Bigr)
\minCZ_{mb}.
\]
CZ conjugation gives
\[
\minE_m(K)=J_{zx}X_b+J_{zy}Y_b.
\]
Hence $J_{zx}=J_{zy}=0$. Invertibility then requires $J_{zz}\ne0$
and an invertible upper-left $2\times2$ submatrix. But the three
nonidentity coefficients of $K$ on $m$ are now
\begin{align*}
K_x&=J_{xz}I_b+J_{yy}X_b-J_{yx}Y_b,\\
K_y&=J_{yz}I_b-J_{xy}X_b+J_{xx}Y_b,\\
K_z&=J_{zz}Z_b.
\end{align*}
They are linearly independent, contradicting the dimension bound.

\subsection{Alternating words with an active middle qubit}
Write the path as $a$--$b$--$c$, where $a$ is inactive and $b,c$
are active, and put $A=ab$, $B=bc$. After the initial local layer,
the source in Eq.~\eqref{SM-eq:min-square} is
$I_a\otimes P_b\otimes P_c$, where $P_b,P_c$ are Pauli operators
along arbitrary unit axes.

\subsubsection{ABAB: the inactive leaf participates in the first CZ}
After the first $A$, the operator has the form
\[
I_a\otimes A_0+Z_a\otimes B_0,\qquad
B_0=P_{b,\perp}\otimes P_c,\qquad B_0^2=\beta^2I_{bc},
\]
where $P_{b,\perp}$ is the part transverse to the first CZ axis $Z_b$.
Until the next $A$, operations on $bc$ conjugate $A_0,B_0$
simultaneously, while a local rotation of $a$ rotates $Z_a$.
Immediately before the third gate we may therefore write
\[
F=I_a\otimes A+(cZ_a+sX_a)\otimes B,\qquad
B^2=\beta^2I,\quad c^2+s^2=1.
\]
The transverse direction can be chosen as $X_a$ because rotations
around $Z_a$ commute with this CZ. Define
\[
B_+=\frac{B+Z_bBZ_b}{2},\qquad
B_-=\frac{B-Z_bBZ_b}{2}.
\]
After this CZ, the $X_a,Y_a$ coefficients are
\[
F_x=sZ_bB_+,\qquad F_y=-\ii sZ_bB_-.
\]
If $s\beta\ne0$, these two matrices have no nonzero common kernel:
$F_xv=F_yv=0$ implies $B_+v=B_-v=0$, hence $Bv=0$, contradicting
$B^2=\beta^2I$. Consequently, the common kernel of all three nonidentity coefficients
is also trivial. The last gate $B$ acts only on $bc$ and preserves
its dimension; local rotations likewise preserve the dimension of
the common kernel of the coefficient triple. This conflicts with the singlet kernel
of the coefficients of $\mathcal R$.

If $s=0$ or $\beta=0$, only the directions $I_a,Z_a$ remain after
the third gate. The operator Schmidt rank is then at most two,
whereas it equals four for $\mathcal R$. This also excludes the
degenerate cases.

\subsubsection{BABA: the inactive leaf participates in the last CZ}
Before the first $A$, the source still has the form
$I_a\otimes R_{bc}$. After this $A$,
\[
F=I_a\otimes A_0+Z_a\otimes B_0,\qquad
B_0=\frac{R-Z_bRZ_b}{2},\qquad \Tr B_0=0.
\]
The next $B$ and the local layers preserve the trace of the
corresponding coefficient. Thus, before the last $A$,
\[
F=I_a\otimes A+(cZ_a+sX_a)\otimes B,\qquad \Tr B=0.
\]
Expand
$B=I_b\otimes B_I+X_b\otimes B_x+Y_b\otimes B_y+Z_b\otimes B_z$.
At the last CZ, a contribution to the $X_a$ row of the $a,b$
pair-interaction matrix can arise only from
$sX_a\otimes I_b\otimes B_I$, which maps to
$sX_aZ_b\otimes B_I$. Its coefficient after tracing out $c$ is
\[
\frac{s}{2}\Tr B_I=\frac{s}{4}\Tr B=0.
\]
The terms $X_aX_b$ and $X_aY_b$ map to the $Y_a$ direction,
while $X_aZ_b$ becomes a local $X_a$ term. Terms with $I_a,Z_a$
create no $X_a$ direction. The entire $X_a$ row of the interaction
matrix therefore vanishes. Its rank is at most two, contradicting
rank three for $\mathcal R$.

These two arguments cover both active-middle placements $m=1,2$;
the preceding subsection covers $m=3$.

\subsection{Nested words: ABBA and BAAB}
We return to $UTU^\dagger=H_0$. Let the path be
$a$--$b$--$c$, with $A=ab$, $B=bc$. We allow the two middle CZ
gates and their local layers to be replaced by an arbitrary unitary
on $bc$.

Assigning $a,b,c$ to the logical labels changes the cyclic $H_0$
by at most an overall sign. The conditions used below are invariant
under $H_0\mapsto-H_0$; all placements of the two active fields are
handled explicitly.

\subsubsection{An invariant after the first outer CZ}
For the coefficients $F_i$ on $a$, define
\[
t_i=\tfrac14\Tr F_i,\qquad M_i=F_i-t_iI,\qquad
M_v=\sum_i v_iM_i.
\]
Write $t=(t_x,t_y,t_z)$ for the vector of trace components.
After the first CZ, each $M_i$ is a traceless matrix on $b$ tensored
with $I_c$. Therefore
\begin{equation}\label{SM-eq:min-cliff}
M_v^2\in\mathbb RI\qquad\text{for every }v\in\mathbb R^3.
\end{equation}
An arbitrary unitary on $bc$ and local rotations of $a$ preserve
this property. Equivalently, every anticommutator $\{M_i,M_j\}$
is scalar.

CZ is locally equivalent to
\[
C(n,m)=\exp[-\ii\pi(n\cdot\minsig_a)(m\cdot\minsig_b)/4],
\qquad |n|=|m|=1.
\]
After absorbing final local layers into the axes $n,m$, a necessary
condition is that the coefficients of
$F=C(n,m)H_0C(n,m)^\dagger$ obey Eq.~\eqref{SM-eq:min-cliff}.
Changing $n\mapsto-n$ covers the opposite sign of the exponent.
External local conjugations preserve the properties being tested.

\subsubsection{Allowed axes of the last CZ}
Let $x,y,z$ be the standard unit vectors in $\mathbb R^3$, and set
$Jv=-z\times v$ and $d_v=z\times v$. Then
\[
H_0=I_a\otimes(\minsig_b^TJ\minsig_c)
+\sum_i\sigma_a^i\otimes D_i,\qquad
D_v=d_v\cdot(\minsig_b-\minsig_c).
\]
For $G=(n\cdot\minsig_a)(m\cdot\minsig_b)$, conjugation by $C$
preserves the part commuting with $G$ and maps an anticommuting
part $O$ to $-\ii GO$. The coefficient along $n$ is already
centered and equals
\begin{equation}\label{SM-eq:min-Mn}
M_n=s(m\cdot\minsig_b)-d_n\cdot\minsig_c
+\minsig_b^TK\minsig_c,\qquad
s=m\cdot d_n,\quad K=m_zI-zm^T.
\end{equation}
The pair term follows from $[m]_\times J=m_zI-zm^T$, where
$[m]_\times v=m\times v$.

For any real matrix $K$, Pauli multiplication gives
\[
(\minsig_b^TK\minsig_c)^2=\|K\|_F^2I
-2\sum_{ij}\mincof(K)_{ij}\sigma_b^i\sigma_c^j,
\]
where $\mincof(K)$ is the cofactor matrix. Since $m^TK=0$, the
nonscalar terms in the square of Eq.~\eqref{SM-eq:min-Mn} are
\[
-2(Kd_n)\cdot\minsig_b
-2\sum_{ij}[\mincof(K)+smd_n^T]_{ij}\sigma_b^i\sigma_c^j.
\]
It follows that
\begin{equation}\label{SM-eq:min-constraints}
Kd_n=0,\qquad \mincof(K)+smd_n^T=0.
\end{equation}
If $m_z\ne0$, $K$ has rank two and kernel $\minspan(z)$.
As $d_n\perp z$, the first condition implies $d_n=0$, while the
second requires $\mincof(K)=0$, impossible at rank two. Hence
$m_z=0$. Then $K=-zm^T$, and the first condition gives $s=0$.
Thus $n\in\minspan(m,z)$.

A common rotation around $z$ leaves $H_0$ invariant, so we may set
\[
m=x,\quad n=\alpha x+\beta z,\quad
u=\beta x-\alpha z,\quad v=y,\qquad \alpha^2+\beta^2=1.
\]
The same conjugation rule yields
\begin{align}
M_n&=-\alpha Y_c-Z_bX_c,\nonumber\\
M_u&=\beta Y_b-X_bX_c,\label{SM-eq:min-centered}\\
M_v&=-\beta X_bY_c.\nonumber
\end{align}
In particular,
\[
\{M_n,M_v\}=2\alpha\beta X_b-2\beta Y_bZ_c.
\]
For this to be scalar, $\beta=0$ is necessary. Therefore the last
CZ axes must be parallel or antiparallel in the equatorial plane.
Since $C^\dagger=\ii GC$, the Hamiltonians obtained for the two signs
are related by conjugation with the local operator $G$.
It suffices to consider $n=m=x$.

\subsubsection{Contradiction for the remaining axes}
For $n=m=x$, direct conjugation gives
\begin{equation}\label{SM-eq:min-forced}
F=I_a\otimes Q_*+X_a\otimes R+Z_a\otimes(-I+S),
\end{equation}
where
\[
Q_*=Z_b+X_bY_c,\qquad R=-Y_c-Z_bX_c,\qquad S=X_bX_c.
\]
Thus $t\ne0$, and the centered coefficients span a two-dimensional
space. Define
\[
\mathcal N(F):=\{F_v:\ v\cdot t=0\}.
\]
For this operator,
\begin{equation}\label{SM-eq:min-notmember}
\mathcal N(F)=\minspan(R),\qquad F_0=Q_*\notin\mathcal N(F).
\end{equation}
The nonmembership follows, for example, from the disjoint Pauli
supports of the nonzero operators $Q_*,R$. All these properties
are preserved by a local rotation of $a$ and simultaneous unitary
conjugation of the coefficients on $bc$.

Now compare with the operator after the \emph{first} outer CZ.
If $a$ is inactive, no local component on $a$ can be created, so
$t=0$. Hence $a$ must be active. If $b$ is inactive, the only
centered coefficient on $a$ comes from the transverse component of
its initial field, and the centered span has dimension at most one.
Thus $b$ must also be active and $c$ inactive.

Choose the first entangler as $\exp(-\ii\pi Z_aZ_b/4)$ and write
the initial unit fields as $a_\parallel Z_a+a_\perp X_a$ and
$b_\parallel Z_b+b_\perp X_b$. Their transverse axes can be chosen
independently. After the entangler,
\begin{align}
F_0&=\sqrt3 b_\parallel Z_b,\nonumber\\
F_z&=\sqrt3(a_\parallel I+b_\perp Y_b),\label{SM-eq:min-first}\\
F_y&=\sqrt3 a_\perp Z_b,\qquad F_x=0.\nonumber
\end{align}
The condition $t\ne0$ requires $a_\parallel\ne0$, and the
two-dimensional centered span requires $a_\perp b_\perp\ne0$.
Consequently,
\[
\mathcal N(F)=\minspan(Z_b),\qquad F_0\in\mathcal N(F).
\]
The middle block preserves this inclusion, contradicting
Eq.~\eqref{SM-eq:min-notmember}. This excludes $ABBA$. Exchanging
the edge names gives the same conclusion for $BAAB$.

\subsection{Attainability and scope}
For the chain $1$--$2$--$3$, set $D_{12}=D_{23}=D_{31}=1$ in
the circuit and angle formulas of Sec. 5 of the main text.
The angles become
\[
(\beta_1,\beta_2,\beta_3,\beta_4)
=(3\pi/4,-\pi/4,\pi+\theta,-\pi-\theta),\qquad
\theta=\operatorname{atan2}(1,\sqrt2).
\]
The direct Pauli conjugation established in Appendix A.4 gives
$WH_0W^\dagger=T$. This proves the five-CNOT upper bound for
$m=2$ and completes Theorem~\ref{SM-thm:path-minimality}.

Every exclusion above concerns the conjugation condition
$UTU^\dagger=H_0$ or its necessary consequence
Eq.~\eqref{SM-eq:min-square}; no particular diagonalizer was fixed.
The proof therefore allows the full freedom $W\mapsto VW$ with
$[V,T]=0$, including arbitrary mixing within the eigenspaces of
dimensions $2,4,2$. A nonlocal $V$ is not supplied as a free gate:
only the final circuit within the specified CZ budget is considered.

Relabeling in the proof means simultaneously relabeling $H_0,T$
and the path edges, not allowing free physical output permutations.
Active and inactive middle qubits were treated separately, covering
all three original placements.

The same count holds when all three couplings equal any real $D\ne0$.
For $D>0$ the equation reduces by scaling to the unit-coupling case.
For $D<0$ it gives $WH_0W^\dagger=-T$; left multiplication by the
local operator $X_1X_2$ changes $-T$ to $T$ without adding CZ gates.

The proven minimum concerns the diagonalizer $W$, not the full
evolution $\exp(-\ii tH_0)$. The lower bound is established for equal
nonzero couplings. Other choices of $D_{ij}$, including degenerate
cases, require a separate analysis.

\FloatBarrier

\section{Experimental protocol and additional measurements}
\label{SM-sec:sup-hardware}

Three datasets serve complementary purposes. The step-count comparison
measures excitation dynamics and transverse magnetization at several
step counts of the symmetric second-order Trotter formula (Strang);
a separate repeat checks reproducibility at selected
times. A third experiment compares the exact circuits with Strang $r=8$
for transverse magnetization and six-input state transfer in one
processor execution. The datasets remain separate. The two acquisition
rounds within each experiment are used to estimate its observables.

\subsection{Processor and measurement protocol}

The circuits were executed on the 156-qubit superconducting quantum
processor IBM Kingston (\texttt{ibm\_kingston}) using the native gate
set $\{R_z,\sqrt{X},X,\mathrm{CZ}\}$.
Five physical chains, denoted A--E, implement the logical chain
$1$--$2$--$3$, with two-qubit gates applied only to adjacent pairs.
The chains were selected in a preceding control experiment and fixed
before the dynamical measurements and were measured simultaneously.
The main-text amplitude and phase comparison uses chain C in both
experiments. This chain had the lowest mean calibrated CZ error among
the five chains in the original calibration snapshot used for the
step-count comparison, and the same chain is retained for the later
illustration; no claim of the best measured performance is implied.
Five-chain summaries of the transverse magnetization and state transfer
are reported separately below. Physical assignments and calibration
records are preserved in the research archive.
An additional three-qubit register, prepared in $|000\rangle$ and measured
in $Z$, served only as a preparation/readout diagnostic.
The analysis uses the logical bit order $|x_1x_2x_3\rangle$, with $x_1$
the most significant bit.

The experiments use the Pauli convention and Hamiltonian of
Eq.~\eqref{SM-eq:sup-triangle}. The excitation-dynamics example has $J=0$,
$D_{12}=D_{23}=D_{31}=1$, and input $|100\rangle$.
The exchange--DM example has $J=1$,
$(D_{12},D_{23},D_{31})=(1/2,1,1/3)$, and input $|+00\rangle$;
these parameters satisfy Eq.~\eqref{SM-eq:sup-phase-condition}.
All time labels refer to simulated evolution time, with the corresponding
coupling scale set to one, rather than to the duration of a physical
circuit.

The transverse-magnetization and directed-transfer experiment was
performed on 17 September 2026. It compared the exact circuits with
Strang at $r=8$ in one processor execution containing two acquisition
rounds of 1024 shots per setting. The rounds give 2048 shots per chain,
method, input, time, and measurement basis where applicable; they are
not independent calibration campaigns. The transverse components were
measured on the 25-point grid $t=k/12$, $k=0,\ldots,24$, in $XXX$ and $YYY$.
The transfer experiment used six input states, both signs of the DM
coupling, and measurements in $XXX$, $YYY$, and $ZZZ$, as detailed below.
These data are analyzed separately from the earlier experiments.

The step-count comparison experiment was performed on 16 September 2026. The pure DM grid consists
of $t=k/12$, $k=0,\ldots,24$, together with $T/3$, $2T/3$, and
$T$, where $T=\pi/\sqrt3$: 28 times in total, measured in $ZZZ$.
The exchange--DM grid contains the 25 times $t=k/12$, measured separately
in $XXX$, $YYY$, and $ZZZ$. Each setting was acquired twice with
1024 shots per repetition; the reported per-chain estimates combine these
two repetitions, giving 2048 shots per time, method, and basis.

An independent repeat was performed on 17 September 2026, using the same
native gate sequences, physical assignments, and measurement conventions,
without recompiling the selected circuits. It included both simultaneous
measurements and single-chain controls. For pure DM, the repeat uses
the exact circuit and Strang with $r=8$, at $t=k/4$, $k=0,\ldots,8$,
plus $T/3$, $2T/3$, and $T$ (12 times). For exchange--DM it uses the exact
circuit and Strang with $r=16$, at $t=k/6$, $k=0,\ldots,12$, in $XXX$
and $YYY$, with additional $ZZZ$ measurements at $t=0,1,2$.
The repeat again provides two sets of 1024 shots per setting.
The subset was specified without selecting points by their measured
agreement with theory. Submission order was shuffled.
All reported experiments used qubit initialization, a repetition delay of
$250\,\mu\mathrm{s}$, and classified readout, with dynamical decoupling and
gate/measurement twirling disabled.

\subsection{Implemented circuits and observables}

The exact constructions use 10 CZ gates for pure DM and 16 CZ gates for
the phase-constrained exchange--DM example. Both methods retain their
full sequences of entangling gates, including at $t=0$ and at the return time. The baseline is the same
fixed symmetric product in all runs. Writing
$E_{ij}(s)=\exp(-\ii s h_{ij})$ for the corresponding bond term in
Eq.~\eqref{SM-eq:sup-triangle}, and $\delta=t/r$, one step is applied in
the chronological order
\begin{equation}
E_{31}(\delta/2),\quad E_{23}(\delta/2),\quad
E_{12}(\delta),\quad E_{23}(\delta/2),\quad E_{31}(\delta/2).
\label{SM-eq:sup-hw-strang-order}
\end{equation}
Adjacent end factors are merged across steps. Routing to the chain uses
analytic combination of neighboring pair operations with SWAP operations,
and restores the initial logical assignment. Thus the comparison includes
local simplifications of the routed baseline. The CZ counts in
Table~\ref{SM-tab:sup-hw-main} describe these executed circuits; they are not
global lower bounds for all compilations of the product formula.
Across the experiments the tested step counts are $r=1,4,8$ for pure DM
and $r=4,8,16$ for exchange--DM. For $r=8$, the executed Strang circuits
contain 88 CZ gates for pure DM and 105 CZ gates for exchange--DM.

For pure DM we use the difference of the excitation probabilities
\begin{equation}
M(t)=P_2(t)-P_3(t)
     =\frac{\langle Z_3(t)\rangle-\langle Z_2(t)\rangle}{2},
\qquad P_i(t)=\left\langle\frac{I-Z_i}{2}\right\rangle.
\label{SM-eq:sup-hw-transfer}
\end{equation}
Here $P_i$ is the probability that qubit $i$ is excited, irrespective
of the other two qubits. Every measured bit string contributes according
to its second and third bits; no restriction to the one-excitation sector
is imposed.
In particular, $M$ need not equal the exclusive difference
$p_{010}-p_{001}$ in noisy data. For exchange--DM the observables are
$\langle X_1(t)\rangle$ and $\langle Y_1(t)\rangle$, obtained from
separate measurement ensembles. The displayed data are raw estimates:
no readout inversion, postselection, or fitted phase/amplitude correction
is applied.

For the earlier step-count comparison and its independent repeat,
counting error bars for a single-shot observable $w$ are
$\mathrm{SE}=\sqrt{(\langle w^2\rangle-\langle w\rangle^2)/(N-1)}$,
with empirical moments and $N=2048$. For the population difference $M(t)$,
$w=(z_3-z_2)/2$ is evaluated on each measured bit string, so the
covariance of the two populations is included. These error bars do not
include drift or other systematic effects.

For the transverse-magnetization and transfer experiment, counting
variances are computed within each acquisition round. For a five-chain mean, the single-shot value is
$u_{a,n}=\frac15\sum_{k=1}^{5}w_{k,a,n}$, where
$a=1,2$ labels the round and $n=1,\ldots,1024$ labels simultaneous
outcomes. If $\bar u_a$ and $s_a^2$ are the sample mean and unbiased sample
variance within a round, then
\begin{equation}
\bar w=\frac{\bar u_1+\bar u_2}{2},\qquad
\mathrm{SE}^2(\bar w)=\frac{s_1^2+s_2^2}{4\times1024}.
\label{SM-eq:sup-hw-shot-mean}
\end{equation}
Averaging the simultaneous outcomes before estimating the variance
preserves covariance between chains. Per-chain error bars use the same
formula without the average over chains. These counting errors do not
include chain-to-chain dispersion, calibration drift, or systematic
preparation and readout errors.

For each physical chain and each method, the errors over its fixed time
grid are
\begin{align}
\epsilon_M&=\left[\frac{1}{N_t}\sum_t
 \bigl(M_{\rm meas}(t)-M_H(t)\bigr)^2\right]^{1/2},
\label{SM-eq:sup-hw-rmse-pure}\\
\epsilon_\perp&=\left[\frac{1}{N_t}\sum_t
 \left\{\bigl(X_{1,\rm meas}(t)-X_{1,H}(t)\bigr)^2+
        \bigl(Y_{1,\rm meas}(t)-Y_{1,H}(t)\bigr)^2\right\}\right]^{1/2}.
\label{SM-eq:sup-hw-rmse-full}
\end{align}
Here $H$ denotes exact Hamiltonian evolution. Equivalently,
$\epsilon_\perp$ is the root mean square of
$|C_{1,\rm meas}-C_{1,H}|$, with $C_1=\langle X_1\rangle+\ii\langle Y_1\rangle$; there is no additional
factor $1/\sqrt2$. The noiseless discretization error is defined by the
same formulas with measured means replaced by the means of the implemented
ideal circuit. These are observable errors, not state or process
infidelities.

\subsection{Transverse magnetization: exact circuit versus Strang with eight steps}
\label{SM-sec:sup-hw-coherence-r8}

For the exchange--DM model and initial state $|+00\rangle$, the exact
16-CZ circuit and the 105-CZ Strang circuit were compared in the same
experiment under the protocol above. Table~\ref{SM-tab:sup-hw-coherence-r8} summarizes the
pointwise means over the five chains; the main-text amplitude and phase
curves use chain C alone. For $O=X_1,Y_1$, the error of a five-chain mean
curve is
$\epsilon_{\bar O}=[N_t^{-1}\sum_t(\bar O_{\rm meas}(t)-O_H(t))^2]^{1/2}$,
where $N_t=25$ and $\bar O_{\rm meas}=\frac15\sum_k O_{k,\rm meas}$.
It is distinct from the mean or median of the five individual RMSEs.

\begin{table}[htbp]
\centering\small
\begin{tabular}{@{}lccc@{}}
\hline
Component & Exact, measured & Strang $r=8$, measured & Strang $r=8$, noiseless\\
\hline
$\langle X_1\rangle$ & $0.04953$ & $0.16448$ & $0.03365$\\
$\langle Y_1\rangle$ & $0.03571$ & $0.21081$ & $0.03921$\\
\hline
\end{tabular}
\caption{RMSE of the five-chain mean transverse-magnetization curves.
The exact construction has zero ideal discretization error. All 25 times,
including $t=0$, enter each measured error.}
\label{SM-tab:sup-hw-coherence-r8}
\end{table}

\begin{figure}[p]
\centering
\includegraphics[width=\linewidth,height=0.78\textheight,keepaspectratio]{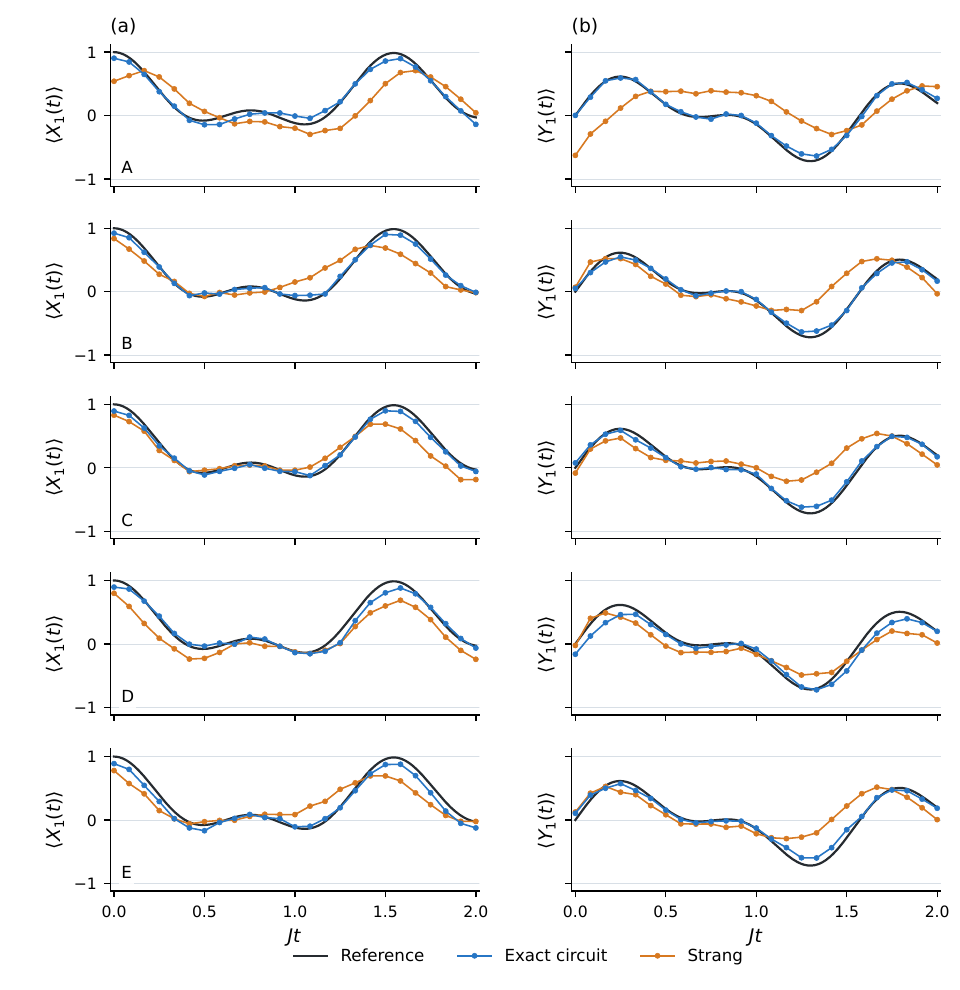}
\caption{Transverse-magnetization dynamics with Strang $r=8$ on all
five chains A--E (rows). Columns show $\langle X_1(t)\rangle$ and
$\langle Y_1(t)\rangle$. Black curves denote exact Hamiltonian evolution;
blue and orange points show the exact 16-CZ circuit and Strang $r=8$
with 105 CZ gates. Each point combines two rounds of 1024 shots; error
bars indicate one counting standard error. All times, including $t=0$,
are retained, with no readout correction or postselection.}
\label{SM-fig:sup-hw-coherence-r8}
\end{figure}

The advantage is also present on every individual chain for each
component, both after combining the rounds and in each round separately
(Fig.~\ref{SM-fig:sup-hw-coherence-r8}). Excluding $t=0$ as a diagnostic
check leaves this ordering unchanged; the reported results include that
point. Thus the advantage is not created by cancellation in the
five-chain mean or by selecting favorable times. The comparison refers
to the fixed implementation of Strang used here.

\subsection{Amplitude and phase on a fixed chain}
\label{SM-sec:sup-hw-amplitude-phase}

The main-text figure compares the same chain C in two separate
experiments: Strang $r=8$ on 17 September and Strang $r=16$ on
16 September 2026. Each is compared with its own exact-circuit control
measured in the same experiment. The exact circuit uses 16 CZ gates;
the two Strang circuits use 105 and 201 CZ, respectively. Each curve
contains all 25 times on $0\leq Jt\leq2$, including $t=0$.
No observations or uncertainties are pooled between experiments.

We first aggregate the $X$ and $Y$ measurements separately; phases
are not averaged across rounds or chains. We form $C_1=x+\ii y$, where $x=\langle X_1(t)\rangle$ and
$y=\langle Y_1(t)\rangle$, and plot its modulus $R=|C_1|$ and principal
argument $\varphi_1=\arg C_1\in[-\pi,\pi]$ (in degrees in the figure).
Equations~\eqref{SM-eq:sup-amplitude} and~\eqref{SM-eq:sup-magnetization} give
$C_{1,H}(t)=a_1(t)e^{4\ii t}$ and $R_H=|a_1(t)|$ for $J=1$.
The real amplitude $a_1$ vanishes at $Jt\simeq0.618012$ and $0.886346$.
On crossing either zero, the sign change of $a_1$ adds $\pi$ to the
phase; at the zero the phase is undefined. Neither zero lies on the measured
time grid. Gray shading marks $R_H<0.25$, where phase errors relative
to the ideal signal are poorly conditioned; this is not a data filter.

The $X$ and $Y$ estimates come from independent measurement ensembles.
For their counting standard errors $s_x,s_y$, first-order uncertainty
propagation gives the modulus and phase standard errors, the latter in radians:
\begin{equation}
s_R^2\simeq\frac{x^2s_x^2+y^2s_y^2}{R^2},\qquad
s_{\varphi}^2\simeq\frac{y^2s_x^2+x^2s_y^2}{R^4}.
\label{SM-eq:sup-hw-phase-se}
\end{equation}
Open markers flag measured $R/\sqrt{s_x^2+s_y^2}<5$. They remain in the
figure without these linear error bars or connecting phase segments,
since the approximation becomes unreliable at weak signal. The
amplitude line retains all points; phase lines also break at the
principal-argument boundary, and the ideal phase line also at signal
zeros. The modulus is not corrected for its
positive statistical bias near zero. No phase alignment, unwrapping,
zero-time subtraction, or amplitude rescaling is applied.

As an illustration away from the zeros, all four curves have measured
signal-to-noise ratio above 10 at the seven sampled times in
$1.25\leq Jt\leq1.75$, where the ideal modulus is at least $0.722$.
The absolute wrapped phase error
$\Delta\varphi=|\arg(C_{1,\rm meas}C_{1,H}^{*})|$ is
$28.47^\circ$--$46.32^\circ$ for $r=8$, versus
$1.58^\circ$--$4.85^\circ$ for its exact control, and
$55.09^\circ$--$109.56^\circ$ for $r=16$, versus
$2.50^\circ$--$4.47^\circ$ for its exact control.
The ideal Strang errors in the same interval are only
$1.16^\circ$--$3.04^\circ$ and $0.30^\circ$--$0.97^\circ$.
The phase distortions substantially exceed the ideal discretization error.
Because these measurements were made in separate experiments, the
effects of the step count and changes in processor conditions cannot be
disentangled.

\subsection{Directional transfer of six input states}
\label{SM-sec:sup-hw-state-transfer}

For this experiment $J=0$, $D_{12}=D_{23}=D_{31}=\eta$, with
$\eta=\pm1$, and $t_*=\pi/(3\sqrt3)$. The first spin is prepared in
one of the six Pauli eigenstates,
\[
|\pm X\rangle=\frac{|0\rangle\pm|1\rangle}{\sqrt2},\quad
|\pm Y\rangle=\frac{|0\rangle\pm\ii|1\rangle}{\sqrt2},\quad
|{+Z}\rangle=|0\rangle,\quad |{-Z}\rangle=|1\rangle,
\]
while the other two spins
start in $|00\rangle$. For each method and sign of DM, the evolution
circuit is held fixed across all six inputs; only preparation and the
measurement basis change. The ideal target propagator transfers the
state to spin 2 for $\eta=+1$ and to spin 3 for $\eta=-1$, with no
additional relative-phase correction. The exact and Strang $r=8$ circuits use 10 and 88 CZ gates,
respectively.

For each input, three measurement settings are used: $XXX$, $YYY$,
and $ZZZ$. In the $XXX$ setting, all three spins are measured along $X$;
averaging the individual $\pm1$ outcome on spin $q$ gives
$x_q=\langle X_q\rangle$. The components $y_q=\langle Y_q\rangle$
and $z_q=\langle Z_q\rangle$ are obtained analogously. These settings
therefore estimate the Bloch vector $\boldsymbol v_q=(x_q,y_q,z_q)$
of every output spin and suffice for local tomography of its reduced state,
\[
\rho_q=\frac{\Id+x_qX+y_qY+z_qZ}{2}.
\]

The overlap with the prescribed input, whose unit Bloch vector is
$\boldsymbol n$, follows directly from the measured components:
\begin{equation}
F_{\psi,q}=\langle\psi|\rho_q|\psi\rangle
 =\frac{1+\boldsymbol n\cdot\boldsymbol v_q}{2}
 =\frac{1+s\langle\sigma_q^\alpha\rangle}{2}
 \quad\text{for }\boldsymbol n=s\boldsymbol e_\alpha,\ s=\pm1.
\label{SM-eq:sup-hw-transfer-fidelity}
\end{equation}
For example, input $|{+X}\rangle$ requires only $\langle X_q\rangle$;
the other two components are also measured. We evaluate $F_{\psi,q}$
using this linear expression without fitting a density matrix.
The reported transfer fidelity is the arithmetic mean of this state
fidelity over the six inputs and the five preselected chains, evaluated
at the prescribed receiver. It includes state-preparation and measurement
errors. No process fidelity, corrected channel, or fitted phase is
inferred from this quantity. For each input the counting standard error
is half that of the relevant spin component. The six inputs are
measured in separate ensembles. The counting variance of their mean
is therefore the sum of the six variances divided by 36;
the five-chain and two-round averages follow
Eq.~\eqref{SM-eq:sup-hw-shot-mean}.

\begin{table}[htbp]
\centering\small
\setlength{\tabcolsep}{4pt}
\begin{tabular}{@{}lcccc@{}}
\hline
Input & \multicolumn{2}{c}{$1\to2$, $\eta=+1$} & \multicolumn{2}{c}{$1\to3$, $\eta=-1$}\\
 & Exact & Strang $r=8$ & Exact & Strang $r=8$\\
\hline
$|{+X}\rangle$ & $0.9477\pm0.0023$ & $0.8438\pm0.0036$ & $0.9500\pm0.0021$ & $0.8521\pm0.0036$\\
$|{-X}\rangle$ & $0.9401\pm0.0023$ & $0.8286\pm0.0037$ & $0.9447\pm0.0023$ & $0.8432\pm0.0036$\\
$|{+Y}\rangle$ & $0.9557\pm0.0020$ & $0.8449\pm0.0037$ & $0.9507\pm0.0022$ & $0.8411\pm0.0035$\\
$|{-Y}\rangle$ & $0.9475\pm0.0022$ & $0.8286\pm0.0037$ & $0.9502\pm0.0022$ & $0.8313\pm0.0037$\\
$|{+Z}\rangle$ & $0.9643\pm0.0019$ & $0.8754\pm0.0033$ & $0.9549\pm0.0020$ & $0.8813\pm0.0031$\\
$|{-Z}\rangle$ & $0.9529\pm0.0021$ & $0.8088\pm0.0039$ & $0.9507\pm0.0022$ & $0.8246\pm0.0037$\\
\hline
\end{tabular}
\caption{Receiver state fidelity for each input, averaged over five
chains and two acquisition rounds. Errors are one counting standard
error and include the covariance of simultaneously measured chains.}
\label{SM-tab:sup-hw-transfer-inputs}
\end{table}

For transfer $1\to2$, the mean fidelity is $0.95135\pm0.00087$ for the
exact circuit and $0.83835\pm0.00149$ for Strang. For $1\to3$, the
corresponding values are $0.95020\pm0.00088$ and $0.84559\pm0.00145$.
The uncertainties are counting standard errors. Averaging only the four
equatorial inputs, which have equal populations and different relative
phases, gives $0.94773$ and
$0.83647$ for $1\to2$, and $0.94890$ and $0.84192$ for $1\to3$.
Across the individual chains and directions, the six-input averages
range from $0.9400$ to $0.9647$ for the exact circuit and from $0.8233$
to $0.8630$ for Strang. The exact result is higher in all 20 combinations
of chain, direction, and acquisition round.

\begin{figure}[p]
\centering
\includegraphics[width=\linewidth,height=0.78\textheight,keepaspectratio]{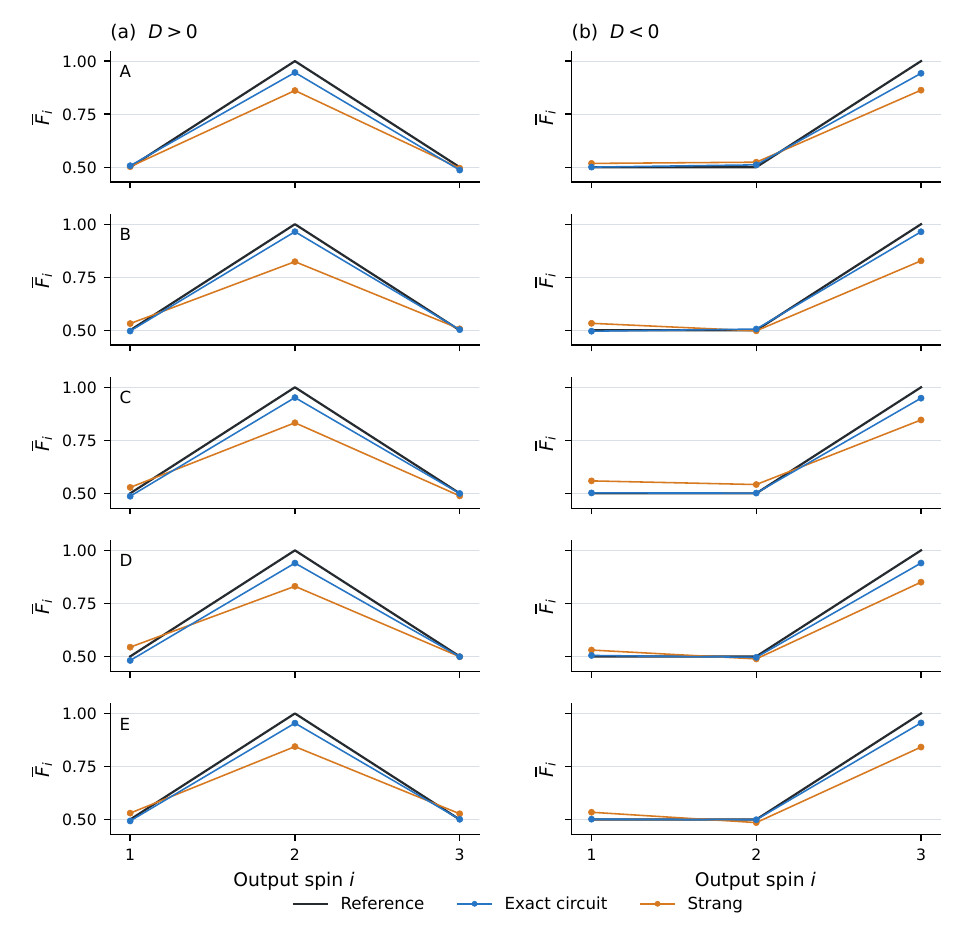}
\caption{Directed state transfer on all five chains A--E (rows).
Columns correspond to positive and negative DM coupling, whose prescribed
receivers are spins 2 and 3. At each output spin the plotted overlap is
averaged over six Pauli input states. Blue and orange points denote the
exact 10-CZ circuit and Strang $r=8$ with 88 CZ gates. Black lines show
the ideal values: unity at the receiver and $1/2$ at the other outputs.
Error bars show one counting standard error from two rounds of 1024
shots per input and basis. Connecting lines indicate the output-spin
positions and do not represent time evolution.}
\label{SM-fig:sup-hw-state-transfer}
\end{figure}

The ideal mean Strang transfer fidelities are $0.99999859$ for $1\to2$
and $0.99999419$ for $1\to3$. Its observed loss of approximately ten percentage points
relative to the exact implementation therefore cannot be attributed to
this discretization error. The shorter circuit gives better transfer
under the tested conditions; this experiment does not separate the
contributions of two-qubit errors, one-qubit errors, and circuit duration.
Figure~\ref{SM-fig:sup-hw-state-transfer} shows the six-input mean overlap at
every output spin for each chain. In the ideal dynamics the two
nonreceiving spins remain in $|0\rangle$, whose overlap averaged over the
six input states is $1/2$. Lines joining the three outputs indicate spin
positions, not a time trajectory.

\subsection{Step-count comparison and independent repeat}

Table~\ref{SM-tab:sup-hw-main} includes every method tested in the step-count comparison experiment.
The summaries are medians and minimum--maximum ranges of five separate
per-chain errors. The range describes variation between these physical
chains; it is not a confidence interval. On every chain the exact circuit
has a smaller time-averaged error than each tested Strang circuit for the
corresponding observable. Increasing the number of Strang steps reduces
the noiseless discretization error, but need not improve the measured
result when the circuit contains more gates. This comparison is restricted
to the tested step counts and implementations.

\begin{table}[htbp]
\centering\small
\renewcommand{\arraystretch}{1.12}
\begin{tabular}{@{}llrcc@{}}
\hline
Example & Method & CZ & \shortstack{Measured RMSE\\median [min, max]} &
\shortstack{Noiseless\\RMSE}\\
\hline
Pure DM & Exact & 10 & $0.0617\ [0.0570,0.0796]$ & $0$\\
 & Strang $r=1$ & 18 & $0.3268\ [0.3202,0.3293]$ & $0.3169$\\
 & Strang $r=4$ & 48 & $0.1264\ [0.1103,0.1288]$ & $0.0145$\\
 & Strang $r=8$ & 88 & $0.1995\ [0.1952,0.2219]$ & $0.0039$\\
\hline
Exchange--DM & Exact & 16 & $0.1042\ [0.0598,0.1245]$ & $0$\\
 & Strang $r=4$ & 57 & $0.2563\ [0.2105,0.3078]$ & $0.1330$\\
 & Strang $r=16$ & 201 & $0.4637\ [0.4055,0.7184]$ & $0.0144$\\
\hline
\end{tabular}
\caption{All five preselected chains in the step-count comparison experiment. Pure DM uses
$\epsilon_M$ over 28 times; exchange--DM uses $\epsilon_\perp$ over
25 times. The noiseless error uses the same observable and grid and is
independent of physical placement. The exact-circuit zeros are analytic
zeros up to numerical roundoff.}
\label{SM-tab:sup-hw-main}
\end{table}

\begin{figure}[p]
\centering
\includegraphics[width=\linewidth,height=0.79\textheight,keepaspectratio]{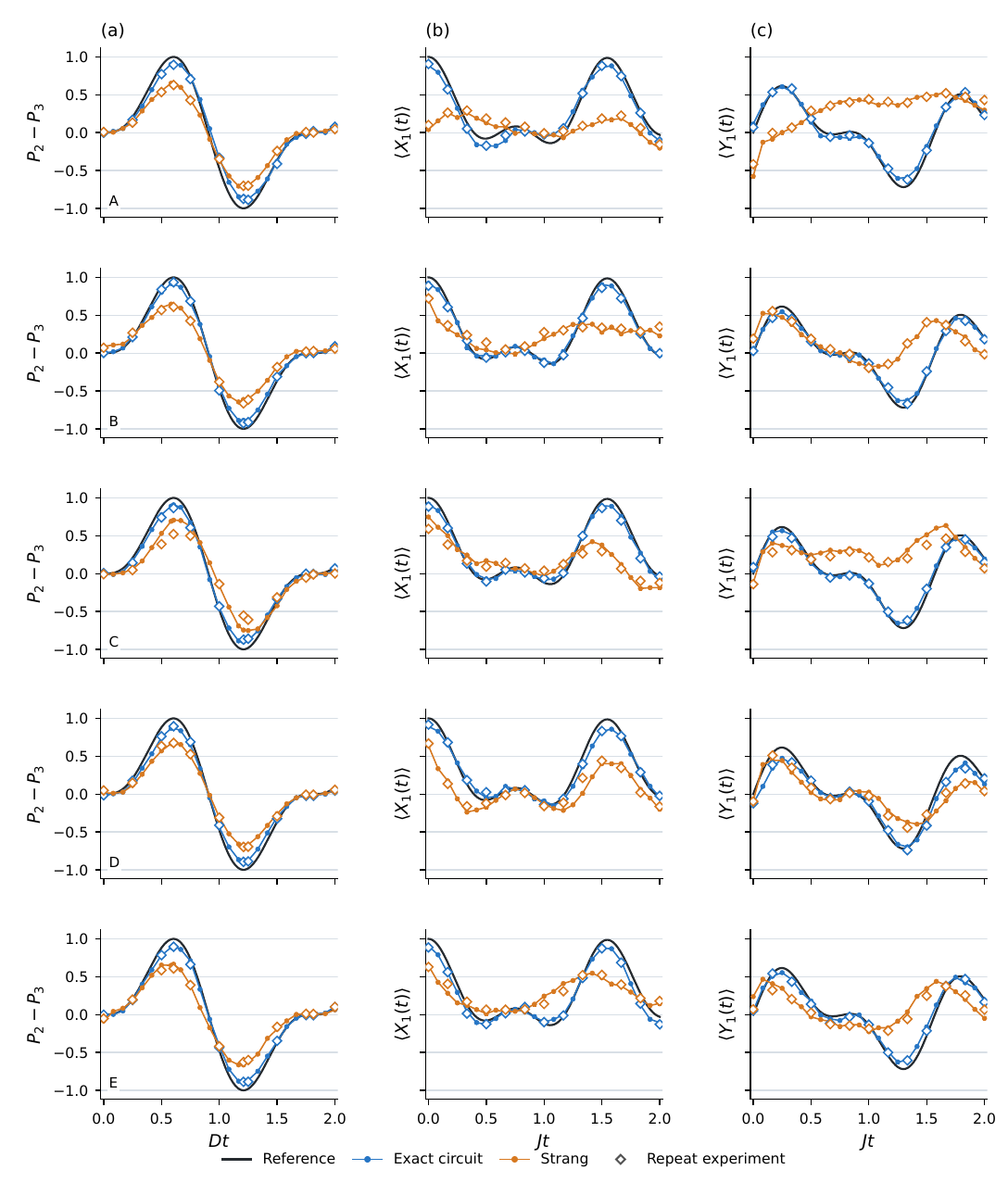}
\caption{Dynamical measurements on all five preselected physical chains.
Rows correspond to chains A--E from top to bottom.
The columns show the excitation-probability difference $P_2-P_3$ for pure DM and the two
transverse components for exchange--DM. Black curves are exact Hamiltonian
evolution. Blue and orange filled circles denote the exact and Strang
measurements in the step-count comparison experiment, respectively; Strang uses $r=8$ for pure DM and $r=16$
for exchange--DM. Thin colored lines connect the step-count comparison points as guides
to the eye; error bars show one counting standard error. Open diamonds
denote the independent repeat, with colors identifying the same methods.
Each point is obtained from 2048 shots on the indicated chain and basis.
The excitation probabilities include every outcome; neither excitation-sector
postselection nor readout mitigation is applied.}
\label{SM-fig:sup-hw-all-paths}
\end{figure}

The repeat is compared with the step-count comparison experiment after restricting the
latter to the same methods and times (Table~\ref{SM-tab:sup-hw-repeat}).
The two datasets are analyzed separately. The lower errors of the
exact circuits persist on each of the five chains in the repeat.
Figure~\ref{SM-fig:sup-hw-all-paths} shows both experiments for every chain,
using $r=8$ for pure DM and $r=16$ for exchange--DM. These curves supplement
the all-method comparison in Table~\ref{SM-tab:sup-hw-main}.

\begin{table}[htbp]
\centering\small
\renewcommand{\arraystretch}{1.12}
\begin{tabular}{@{}llcc@{}}
\hline
Example & Method & \shortstack{Step-count comparison\\common times} & Repeat\\
\hline
Pure DM & Exact & $0.0643\ [0.0630,0.0792]$ & $0.0688\ [0.0430,0.0884]$\\
 & Strang $r=8$ & $0.2102\ [0.1980,0.2372]$ & $0.2323\ [0.1883,0.2916]$\\
Exchange--DM & Exact & $0.1024\ [0.0574,0.1241]$ & $0.0911\ [0.0775,0.1192]$\\
 & Strang $r=16$ & $0.4666\ [0.3964,0.7296]$ & $0.4685\ [0.3538,0.7020]$\\
\hline
\end{tabular}
\caption{Repeat comparison using identical time grids and physical
assignments: 12 pure DM times and 13 exchange--DM times. Entries are
per-chain RMSE medians [minimum, maximum] over the five chains. The
additional $ZZZ$ measurements of the exchange--DM example are not included
in $\epsilon_\perp$.}
\label{SM-tab:sup-hw-repeat}
\end{table}

\subsection{Controls and reproducibility}

The preceding selection experiment used computational-basis preparation
and readout tests, transverse-axis tests, Bell-state tests on adjacent
pairs, and the complete 16-CZ exact circuit at $t=0$.
The same five chains were used throughout, retaining all times including
$t=0$. All five passed the short controls in the
step-count comparison experiment and its independent repeat. On chain E, the full zero-time circuit gave
$\langle X_2\rangle=-0.1719\pm0.0218$ in the step-count comparison experiment and
$-0.1348\pm0.0219$ in the repeat, where the uncertainties are counting
standard errors. The chain remained in both datasets. In the repeat it passed the
predefined tolerance $|\langle X_2\rangle|-2\,\mathrm{SE}<0.1$;
passing this tolerance does not imply that the residual offset vanished.
The additional diagnostic register showed poor preparation/readout
behavior and remained separate from the five dynamical chains.

In the transverse-magnetization and transfer experiment, each round
included computational-basis controls covering eight known bit patterns
and axis controls for the six Pauli eigenstates on all three spins.
Taking the minimum correct joint-outcome probability over these controls
and both rounds separately for each chain gives five minima ranging
from $0.959$ to $0.971$. The separate diagnostic
register remained anomalous: its smallest measured $P_{000}$ in the
dynamical circuits was $0.288$, although it was prepared in $|000\rangle$.
Its diagnostic role was fixed before execution, and it never entered the
five-chain averages; no new chain or time exclusions were made after
viewing the data. The origin of this anomaly is not resolved by these
measurements and remains a limitation of the readout/control assessment.

In the earlier independent repeat, single-chain controls on chains B and C provide limited
checks against simultaneous execution, including the full zero-time
circuit and selected dynamical settings. These controls do not identify
a unique physical error mechanism or establish statistical independence
of the simultaneously measured chains. The comparisons above therefore
make no causal assignment of the deviations to a particular noise source.

The research package\ifdefempty{\DatasetVersionDOI}{}{
(\href{https://doi.org/\DatasetVersionDOI}{doi:\DatasetVersionDOI})}
contains raw samples and counts, QPY files,
native gate lists, physical assignments, calibrations, execution metadata,
and analysis scripts. Preflight checks verify the embedding and
classical-register ordering; the native programs agree with ideal
$8\times8$ operators to a maximum matrix discrepancy below
$1.2\times10^{-12}$. This verifies software consistency, not experimental
fidelity. The independent repeat used the same native gate sequences.
The scripts reproduce observables and tables from the measured outcomes.
For the transverse-magnetization and transfer experiment, an independent
reconstruction checked the stored measurements against the raw execution
results and reproduced ideal probabilities, observables, and counting
errors to numerical precision. The report with numerical residuals is
preserved in the archive.

\FloatBarrier

\FloatBarrier

\section{Coherent-error compensation}
\label{SM-sec:sup-compensation}

These numerical examples use $J=1$ and
$(D_{12},D_{23},D_{31})=(1/2,1,1/3)$, with ordinary Pauli matrices.
The model includes coherent errors only, with no relaxation or dephasing.
Simulated time $t$ is distinct from physical execution time
$s$. The sequential native block has 16 CZ, 36 SX, one X and 57 virtual
$R_z$ gates. Its duration is $23.6\tau_{\rm CZ}$, with
$\tau_{\rm SX}=0.2\tau_{\rm CZ}$, $\tau_{\rm X}=0.4\tau_{\rm CZ}$
and instantaneous $R_z$ rotations. Rectangular control pulses are
generated, up to irrelevant global phases, by
\[
H_{\rm CZ}=\frac{\pi n_i n_j}{\tau_{\rm CZ}},\qquad
H_{\rm SX}=\frac{\pi X_i}{4\tau_{\rm SX}},\qquad
H_{\rm X}=\frac{\pi X_i}{2\tau_{\rm X}},\qquad
n_i=\frac{I-Z_i}{2}.
\]
The constant unwanted Hamiltonian $K$ acts throughout every finite pulse:
each interval is propagated with $H_{\rm ctrl}+K$.

For the isolated triangle, $K=\kappa Z_1Z_2$ with
$\kappa\tau_{\rm CZ}=0.03$.
The scale $\tau_{\rm CZ}=200$~ns chosen for the numerical estimate gives
$\kappa/(2\pi)=23.9$~kHz, SX/X durations of 40/80~ns,
and a block duration of $4.72\,\mu$s.
In the convention $K=(\zeta/4)Z_1Z_2$, the conditional Ramsey shift
is $\zeta/(2\pi)=4\kappa/(2\pi)=95.5$~kHz.
The chosen gate durations and unwanted coupling strength are comparable
in order of magnitude to published values for superconducting circuits
\cite{SM-Hashim2022ZZScale,SM-Wei2022ZZScale}.
The calculation uses rectangular pulses with the generators specified above.
The existing $R_z$ angles are fitted separately at each $t$, with gate
order and durations fixed, by minimizing a phase-independent unitary
distance. This is equivalent to minimizing
$1-F_{\rm pro}$, where
$F_{\rm pro}=|\Tr(U_\triangle^\dagger V)|^2/8^2$ and $V$ is the
implemented unitary. The fit is nonlinear; it neither changes the target
Hamiltonian nor guarantees exact cancellation.

For composition, $H_A$ and $H_B$ act on triangles $(1,2,3)$ and
$(3,4,5)$, respectively, with the same local parameters.
At $t=1$, two symmetric second-order Trotter (Strang) steps for
$H_A+0.2H_B$ give the sequence
\[
A(0.25),\quad B(0.1),\quad A(0.5),\quad B(0.1),\quad A(0.25),
\]
where
$A(u)=e^{-\ii uH_A}$ and similarly for $B$.
Here $K\tau_{\rm CZ}=0.03\sum_{j=1}^4 Z_jZ_{j+1}$.
Each distinct local block is first calibrated against its three-spin
target, including both internal chain bonds; these angles are then frozen
for both comparisons.

Four spectator $X$ pulses are inserted per block: on qubit 4 during $A$
and on qubit 2 during $B$. Their positions minimize the squared Frobenius
norm of the first-order boundary-error integral
$G=\int y(s)U_a^\dagger(s)Z_bU_a(s)\,ds$, subject to
$\int y(s)\,ds=0$. Here $U_a$ includes the active control and internal
drift, $b$ is its boundary spin and $y=\pm1$ changes at each echo.
Selection assumes instantaneous X pulses and searches only boundaries
between native gates, preserving every CZ and SX pulse.
The selected positions, in active time measured in $\tau_{\rm CZ}$,
are $(5,11,16,21.8)$ for $A$ and $(4,12.6,17,20.2)$ for $B$.
Validation includes finite X pulses, with active control paused and all
ZZ terms retained. The complete sequence still has 80 CZ; the added
20 X pulses increase its duration from 118 to $126\tau_{\rm CZ}$.
This restricted search illustrates control-aware error suppression
\cite{SM-Khodjasteh2012,SM-Xie2022}; it does not establish a minimum pulse count
or an efficient search for arbitrary circuits.

The reported five-spin error is
$\epsilon_4=1-\frac14\sum_{k=1}^4
|\langle\psi_k|U^\dagger V|\psi_k\rangle|^2$, with
$U=e^{-\ii(H_A+0.2H_B)}$ and inputs
$|+0000\rangle$, $|10001\rangle$, $|++000\rangle$, and one fixed
complex Haar-random state (seed 20260924).
It decreases from $78.9769\%$ to $8.1138\%$ with echo.
This finite-input average includes the outer splitting error; it is not
a process infidelity. Saved programs, inputs, and replay scripts are available on Zenodo:
\href{https://doi.org/\DatasetVersionDOI}{doi:\DatasetVersionDOI}.

\FloatBarrier

\begingroup\small

\raggedright

\endgroup

\end{document}